\documentclass[onecolumn]{aa}

\usepackage{graphicx}
\usepackage{natbib}
\usepackage{colortbl}
\usepackage{booktabs}
\defcitealias{Bac99a}{BE99}
\defcitealias{Woi02}{Paper~I}

\begin{document}

\title{HST/STIS observations of the RW~Aurigae bipolar jet: mapping the
physical parameters close to the source.
       \thanks{Based on observations made with the NASA/ESA
       {\it Hubble Space Telescope}, obtained at the Space
       Telescope Science Institute, which is operated by the
       Association of Universities for Research in Astronomy, Inc.,
       under NASA contract NAS5-26555.}}

\author{Stanislav Melnikov \inst{1,2} \and
        Jochen Eisl\"offel \inst{1} \and
        Francesca Bacciotti \inst{3} \and
        Jens Woitas \inst{1} \and
        Thomas P. Ray \inst{4}}

\institute{Th\"uringer Landessternwarte Tautenburg, Sternwarte 5,
           D-07778 Tautenburg, Germany  \and
           Ulugh Beg Astronomical Institute, Astronomical str. 33, 100052 Tashkent, Uzbekistan \and
           I.N.A.F. -- Osservatorio Astrofisico di Arcetri, Largo E. Fermi 5,
           I-50125 Firenze, Italy \and
           Dublin Institute for Advanced Studies, 31 Fitzwilliam Place,
           Dublin 2, Ireland}


\date{Received 22 December 2008 / Accepted 20 July 2009}

\abstract{We present the results of new spectral diagnostic investigations applied to high-resolution long-slit spectra obtained with the Hubble Space Telescope Imaging Spectrograph (HST/STIS) of the jet from the T Tauri star RW~Aur.}
{Our primary goal is to determine basic physical parameters (electron density $n_\mathrm{e}$ and electron temperature $T_\mathrm{e}$, hydrogen ionisation fraction $x_\mathrm{e}$, total hydrogen density $n_\mathrm{H}$, radial velocity $v_\mathrm{r}$ and the mass outflow rate $\dot M_\mathrm{j}$) along both the red- and blueshifted lobes of the RW~Aur jet.}
{The input dataset consists of seven long-slit spectra, of 0\farcs1 spatial resolution, taken with the STIS slit parallel to the jet, and stepped across it. We use the Bacciotti \& Eisl\"offel (1999) method to analyse the forbidden doublets [O\,I]\,$\lambda\lambda$6300,6363, [S\,II]\,$\lambda\lambda$\,6716,6731, and [N\,II]\,$\lambda\lambda$\,6548,6583\,\AA\ to extract $n_\mathrm{e}$, $T_\mathrm{e}$, $x_\mathrm{e}$, and $n_\mathrm{H}$.}
{We were able to extract the parameters as far as $3\farcs9$ in the red- and $2\farcs1$ in the blueshifted beam. The electron density at the base of both lobes is close to the critical density for [S\,II] emission but then it decreases gradually with distance from the source. The range of electron temperatures derived for this  jet ($T_\mathrm{e}$=$10^4$--$2\times10^4$\,K) is similar to those generally found in other outflows from young stars. The ionisation fraction $x_\mathrm{e}$ varies between 0.04 and 0.4, increasing within the first few arcseconds and then decreasing in both lobes. The total hydrogen density, derived as  $n_\mathrm{H} = n_\mathrm{e} / x_\mathrm{e}$, is on average $3.2\times10^4\mathrm{cm}^{-3}$ and shows a gradual decrease along the beam. Variations of the above quantities along the jet lobes appear to be correlated with the position of knots. Combining the derived parameters with $v_\mathrm{r}$ measured from the HST spectra and other characteristics available for this jet, we estimate $\dot M_\mathrm{j}$ following two different procedures. The mass-outflow rate $\dot M_\mathrm{j}$ is moderate and \textit{similar in the two lobes, despite the fact that the well-known asymmetry in the radial velocity persists close to the source.} Using the results of the BE diagnostics we find averages along the first 2\farcs1 of both flows (a region presumably not yet affected by interaction with the jet environment) of $2.6\times10^{-9}\,M_\odot$\,yr$^{-1}$ for the red lobe and $2.0\times10^{-9}\,M_\odot$\,yr$^{-1}$ for the blueshifted flow, with an uncertainty of $\pm \log\,M_\odot$=1.6.}
{{The fact that the derived mass outflow rate is similar in the two lobes appears to indicate that the central engine is constrained on the two sides of the system and that the observed asymmetries are due to environmental conditions. Possible suggestions for the origin of the differences are discussed. The RW~Aur jet appears to be the second densest outflow from a T Tauri star studied so far, but its other properties are quite similar to those found in other jets from young stars, suggesting that a common acceleration mechanism operates in these sources.}}


\titlerunning{A HST/STIS study of the RW~Aurigae bipolar jet}
\maketitle

\section{Introduction}
\label{intro}

Highly-collimated jets from young stellar objects (YSOs) have been thought for many years to be an essential ingredient of the star formation process \citep{MF83}. Their origin, ejection, and collimation is, however, still a matter of debate \citep[e.g.][]{Ray07}. Today, high-resolution spectroscopy from space and the ground allows us new insights into the physical conditions of these jets close to their source. This opens up new ways to compare observations with the different magneto-hydrodynamic (MHD) models proposed for the jet launch and acceleration \citep[e.g.][]{Cam90,Koe00,Shu00,Matt03,Fer04}. During the past few years, high resolution studies of several jets from young stars on subarcsecond scales have become available -- e.g \object{HH\,30} \citep{Bac99b,Har07}, \object{DG~Tau} \citep{Bac00,Lav00,Pyo03}, \object{HL~Tau} \citep{Pyo06}, \object{RW~Aur} \citep{Pyo06}, and \object{LkH$\alpha$~233} \citep{Per07,Mel08}. The present study is a continuation of the spectroscopic analysis of RW~Aur performed with the Space Telescope Imaging Spectrograph (STIS) on-board the Hubble Space Telescope (HST), started by \citet[][hereafter Paper~I]{Woi02}, and \citet{Woi05}.

The classical T Tauri star RW~Aur is a double system in the Taurus-Auriga star forming region. The primary, RW~Aur~A (Sp. K1--K4), is one of the optically brightest T Tauri stars with $V=10\fm1$. The secondary, RW~Aur~B ($V=12\fm7$), is located at a projected separation of 1\farcs2 and a position angle of $256^\circ$ with respect to the primary. RW~Aur might be a multiple hierarchical system, although this is not clear. RW~Aur~A is a suspected spectroscopic binary \citep{Pet01}. Another component, which was resolved very close to the secondary \citep{Ghez93}, was probably a false detection \citep{Whi01}.

A bipolar jet, with well-collimated blue- and redshifted lobes emanating from RW~Aur~A (\object{HH\,229}), was found by \citet{Hir94}. Further work traced the emission out to 15\arcsec\ in the redshifted lobe \citep{Dou00} and over 100\arcsec\ in the blueshifted beam \citep{ME98}. Radial velocity measurements of several forbidden lines revealed an interesting  asymmetry: the radial velocity of the blueshifted lobe reaches about $-190$\,km\,s$^{-1}$, while a velocity of only $+100$\,km\,s$^{-1}$ is measured on the red side \citep{Hir94}. Another unusual property of this jet is that for the first 10\arcsec\ the redshifted jet is brighter (in the red [S\,II] doublet) than the blueshifted beam \citep{ME98}, whereas in general blueshifted jets are brighter. Beyond this distance, the redshifted RW~Aur jet strongly drops in brightness, becoming much fainter than the blueshifted flow.

The physical parameters along the RW~Aur jet were first studied by \citet{Bac96}, using spectra taken from the ground at moderate spatial and spectral resolution, and subsequently by \citet{Dou00}, again from the ground but at higher spatial resolution using adaptive optics. These estimates are discussed in detail later in the present paper. A substantial advance in our understanding of this jet came, however, with the observations at high angular and spectral resolution performed with HST/STIS. In this case, seven long-slit spectra of the RW~Aur jet were taken at 0\farcs1 spatial resolution, with the STIS slit parallel to the jet, and stepped every 0\farcs7 across it. In \citet{Woi02} (hereafter Paper I) we presented reconstructed velocity channel maps of the bipolar jet in the forbidden lines, giving  indications of the morphology and kinematics of the jet close to the star and at high resolution. In the second paper of the series, \citet{Woi05}, we reported signatures of rotation in both jet lobes within the first 1\farcs5 from the central source (about 200\,AU at the distance of RW~Aur). Similar signatures of rotation have been discovered in other  jets from T Tauri stars \citep{Bac02a, Cof04, Cof07}. Such observations support the magneto-centrifugal launching scenario for jets \citep[cf][and references therein]{Fer06,Pud07}, which predicts a transport of angular momentum from the disk to the jet by the accretion/ejection `engine'. Recently, however, the disk surrounding RW~Aur has been found to rotate in the opposite sense of the bipolar jet \citep{Cab06}. An explanation for this apparent discrepancy has not yet been found.

In this third paper of the series, we continue the analysis of our high-resolution HST/STIS spectra of the RW~Aur jet, to derive  the physical quantities in the emitting gas from the observed line ratios. Applying specialised diagnostic techniques we estimate the electron density and temperature, the ionisation fraction, the total hydrogen density and the mass outflow rate along the jet. Here we report on such results in detail, and we offer a new analysis of the spatio-kinematical structure of the bipolar jet extracted from the high-resolution position-velocity maps.

The paper is structured as follows: the observations and details of data reduction and analysis are described in Sect.~\ref{obs}. In Section~\ref{results} we illustrate the estimates of the physical quantities along both lobes of the jet. Then in Section~\ref{disc} we contrast the derived jet properties of the opposing lobes, and we compare the estimates with values derived previously and in other T Tauri jets. Finally, we summarize our conclusions in Sect.~\ref{concl}.

\section{Observations and Data Analysis}
\label{obs}

Our long-slit HST/STIS spectra of the RW~Aur jet  were taken on 10 Dec 2000. Observational details and the basic steps of the data reduction have been described in \citetalias{Woi02} and in \citet{Woi05}. Briefly, a slit of width 0\farcs1, length 52\arcsec and parallel to the RW~Aur jet axis (P.A. = $130\degr$, \citealt{Dou00}) was moved across the jet in steps of 0\farcs07 from south-west (slit S1) to north-east (slit S7). Thus, our dataset consists of seven spectra (S1,..,S7) with a spatial sampling  0\farcs05/pixel, and, with the chosen grating (G750M), a spectral sampling of 0.554\,\AA/pixel, corresponding to about 25\,km\,s$^{-1}$/pix in the selected  wavelength range  (6300 -- 6860\,\AA). The basic reduction of the spectra has carried out by the HST/STIS pipeline, and includes flatfielding, cleaning of cosmics and bad pixels, and calibration in surface brightness and wavelength. The spectra include prominent forbidden lines -- [O\,I]\,$\lambda\lambda$6300,6363, [N\,II]\,$\lambda\lambda$6548,6583, [S\,II]\,$\lambda\lambda$6716,6731, in addition to the stellar continuum and strong H${\alpha}$ emission.

In order to trace the jet in the forbidden lines as close as possible to the source, the continuum has to be removed. This is easily done by subtracting the continuum fitted row by row with a high-order polynomial on a sampling region that excludes the emission lines. H$\alpha$ is also subtracted in the same way, as it is not usable for the diagnostics and its high luminosity affects the [NII] lines. Subtraction residuals, however, grow rapidly for positions close to the source, so the area  inside $\pm 0\farcs15$ is excluded from further analysis and it is masked in all the figures presented in this paper.

In order to improve the quality of the extraction of the flux in the lines and to calculate statistical errors, we determined the background from running averages of boxes of size $7\times8$ (170\,km\,s$^{-1}\times$ 0\farcs4) pixels moving along both sides of the lines in each individual slit, and we then subtracted it from the fluxes of the emission lines along the jet lobes.

\begin{figure*}[t]
\centering
\begin{minipage}[c]{15cm}
\includegraphics[width=15cm]{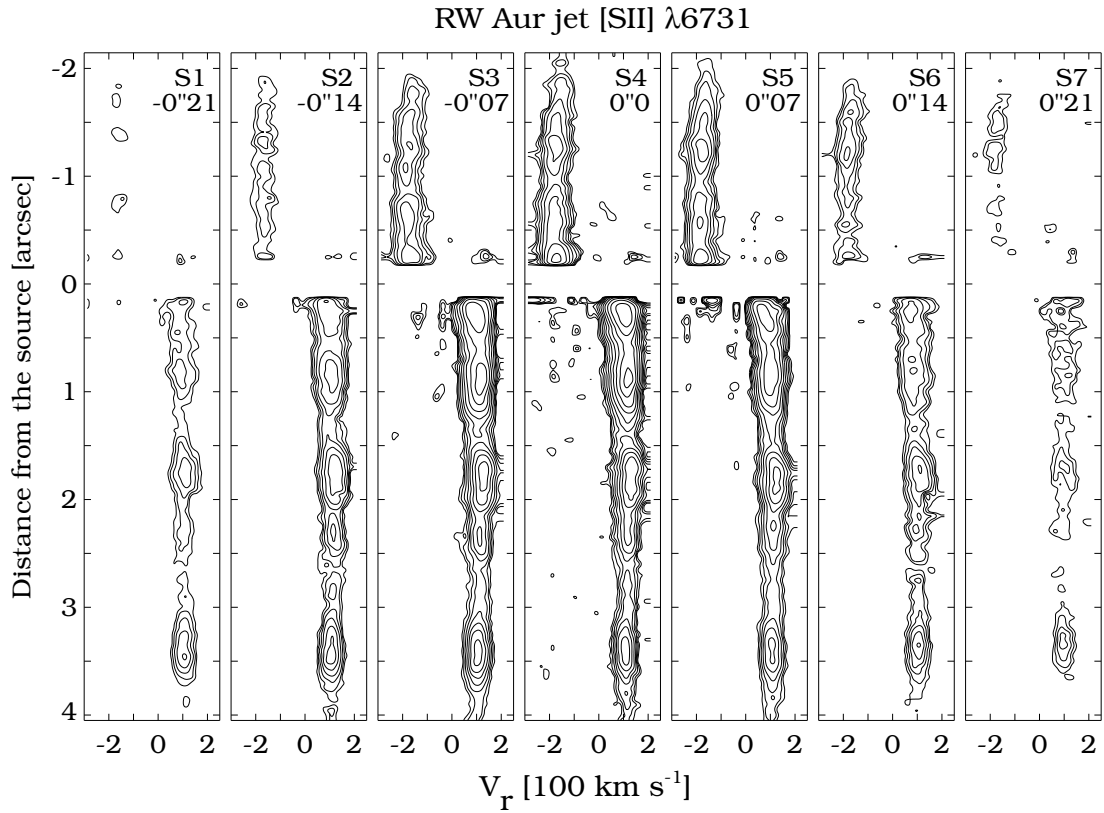}
\end{minipage}
\caption{\label{sevenslits}
Position-Velocity diagrams of the [S\,II]\,$\lambda$6731 line, obtained for the seven positions of the STIS slit stepped across the RW~Aur jet.  The slit was oriented parallel to the jet with P.A. = $130\deg$. All velocities are systemic in this and all subsequent figures (with $v_\star = +23.5$\,km\,s$^{-1}$ in the heliocentric frame). The contours are drawn logarithmically within the same range of surface brightness in all the panels (floor: $-14.7$, i.e. $2\times10^{-15}$erg\,s$^{-1}$\,cm$^{-2}$\,arcsec$^{-2}$\,\AA$^{-1}$; contour step: 0.2 in logarithm).}
\end{figure*}

\begin{figure*}[t]
\centering
\begin{minipage}[c]{15cm}
\includegraphics[width=15cm]{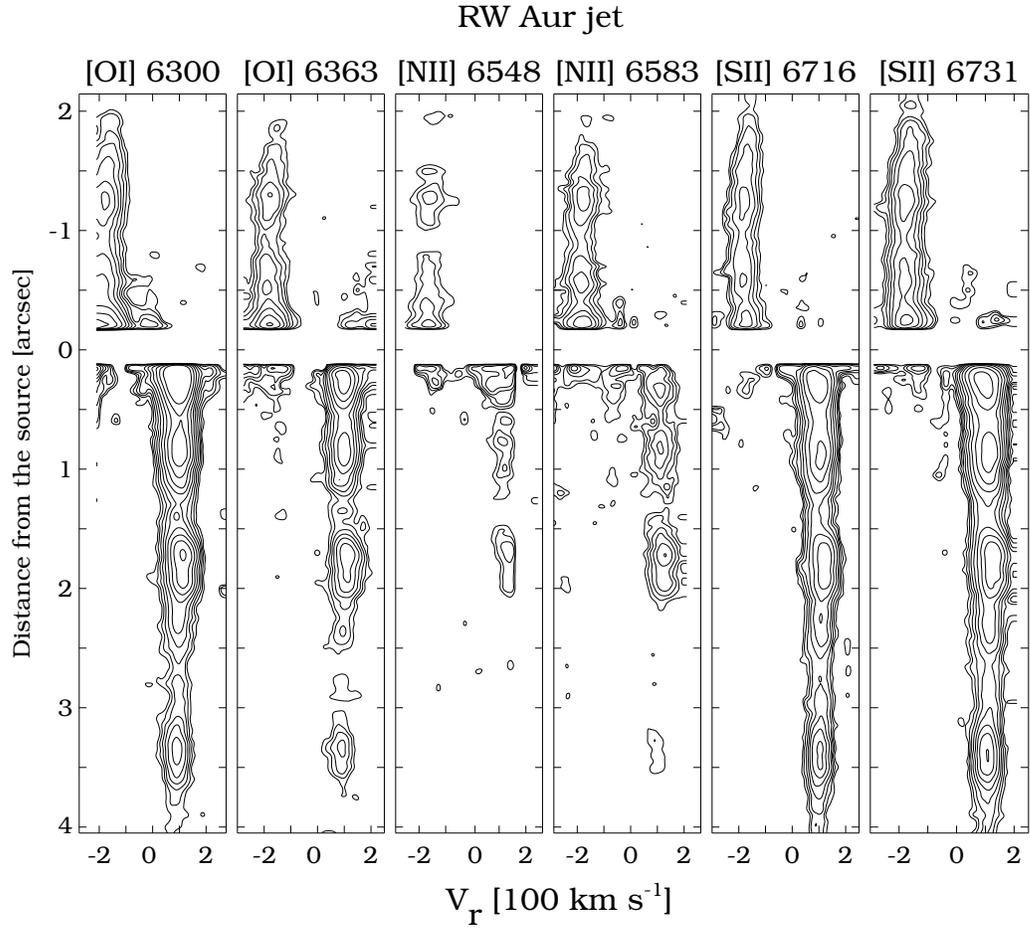}
\end{minipage}
\caption{\label{rvmap}
Position-Velocity diagrams in each of the forbidden lines in the RW~Aur jet, obtained from coadding the spectra in the seven slit positions across the jet. High spatial resolution (0\farcs1) is retained in the direction along the jet. The logarithmic contours are drawn from $-14.2$ ($6.3\times10^{-15}$erg\,s$^{-1}$\,cm$^{-2}$\,arcsec$^{-2}$\,\AA$^{-1}$), with incremental logarithmic step of 0.2.}
\end{figure*}

An example of the cleaned spectra in each of the seven slit positions for the [S\,II]\,$\lambda$6731 line is provided in Fig.~\ref{sevenslits}, in the form of Position-Velocity (PV) diagrams. Similarly, PV diagrams for each emission line, this time obtained from coadding the spectra obtained in all the seven slit positions, are shown in Fig.~\ref{rvmap}. The PV maps approximately span velocities in the range from $-300$ to $+250$\,km\,s$^{-1}$, and are corrected for the heliocentric radial velocity of RW~Aur~A \citep[$+23.5$\,km\,s$^{-1}$,][]{Woi05} . The contours are all scaled to the same range of intensity in each figure.

To apply the diagnostic technique, the resulting pure forbidden flux was further integrated over the line width. This step was dictated by the fact that the emission is concentrated in a highly collimated jet moving in a narrow range of  radial velocities \citepalias[see][]{Woi02}. Therefore, the signal is not strong enough at the lateral borders of the jet, and at the extremes of the velocity range, to allow a meaningful application of the diagnostic technique at the original spatial and spectral resolution. High angular resolution (0\farcs1 on two pixels) is however retained \textit{along} the jet.

To perform the diagnostic analysis, line ratios were then calculated every 0\farcs05 along the jet axis and used to derive the physical quantities of interest (electron density $n_\mathrm{e}$, hydrogen ionisation fraction $x_\mathrm{e}$, electron temperature $T_\mathrm{e}$,  and  total hydrogen density $n_\mathrm{H}$ along the jet). To this aim, we followed the method described in \citet{Bac99a} (hereafter BE99). The so-called {\it BE technique} illustrated in that paper requires as input the flux ratios [S\,II]\,$\lambda$6716/$\lambda$6731, [O\,I]\,$\lambda$(6300+6363)/[S\,II]\,$\lambda$(6716+6363), and [N\,II]\,$\lambda$(6548+6583)/[O\,I]\,$\lambda$(6300+6363) (in the following the latter two ratios will be referred to simply as [O\,I]/[S\,II] and [N\,II]/[O\,I]). The comparison between the ratios calculated theoretically and the ones observed gives estimates of the physical quantities. The ratio of the [S\,II] doublet is used for the determination of  $n_\mathrm{e}$ \citep{Ost89}, up to densities of about $2.5\times10^4$\,cm$^{-3}$, above which the ratio reaches a plateau. The ratio [N\,II]/[O\,I] mostly depends of the ionisation fraction $x_\mathrm{e}$, and [O\,I]/[S\,II] is dependent on both $x_\mathrm{e}$ and $T_\mathrm{e}$.

The errors on the physical quantities extracted with the BE technique originate from the background noise, (calculated from the RMS of the fluxes in running average background boxes) but also come from other sources. Whereas the error of the electron density $n_\mathrm{e}$ depends mainly on  the error of the ratio [S\,II]\,$\lambda$6716/$\lambda$6731 propagated from the noise values, the estimates of temperature $T_\mathrm{e}$ and hydrogen ionisation fraction $x_\mathrm{e}$ depend both on the variations of the ratios [O\,I]/[S\,II] and [N\,II]/[O\,I] respectively and on the uncertainty of the $n_\mathrm{e}$ estimates. Another source of error is the uncertain extinction $A_\mathrm{v}$ not only at the source, but also along the jet. Nevertheless the diagnostic analysis is based on ratios of lines quite close in wavelength, and thus the results are almost independent on dereddening correction, at least for extinction typical of T Tauri stars. In order to evaluate the impact of extinction in the RW Aur case, however, we used a plausible range of $A_\mathrm{v}$ values from $0\fm03$ to $0\fm72$, derived from the study of \citet{Whi01}, to correct forbidden line fluxes. Comparison of the resulting difference with the noise values shows that the former are only a few percent of the S/N errors. In turn, we expect that errors due to variation of extinction along the jet beams are negligible, and therefore we used a constant average value $A_\mathrm{v}= 0\fm4$ everywhere along the jet to find the physical parameters. As we will discuss in Section\,\ref{mdot}, however, extinction variations may play an important role in calculations involving the absolute flux of the lines, as used to estimate the jet mass outflow rate. The results of the BE technique also depend on the adopted abundance of chemical species. \citet{Pod06} showed  that the interstellar gas composition determined for the Orion nebula by \citet{Est04} led to the most reliable diagnostic results for jets in various regions. Thus, we adopted the same abundance set, assuming RW~Aur star formation region has been exposed to the same enrichment processes.

Finally, we note that a fundamental assumption for the application of the BE technique is that the minimum region sampled is homogenous in the physical conditions of the plasma, in such a way that the excitation of the different lines can be well described by a unique electron temperature, electron density and ionisation fraction. As demonstrated in \citetalias{Bac99a}, this condition holds in the cooling region of shock fronts with shock velocities between 30 and 70\,km\,s$^{-1}$, typical of jets. In the present case, however, the line ratios are calculated after coadding the spectra in the transverse direction, which  implies the additional assumption that at each distance from the star the gas conditions are nearly homogenous in the jet transverse direction. In the present case, such an assumption is justified by the fact that the images and spectra of the flow show an extremely well collimated jet with very little dispersion in velocity. This supports a scenario in which the jet flow intersects the internal bow shock fronts in a small area at their apex, in which the front can be thought to be perpendicular to the flow and cover the jet transverse section with similar shock strength. In turn, this indicates  homogenous excitation conditions of the gas across the transverse jet section.

In order to calculate the mass outflow rate in the jet one needs an accurate determination of the jet velocity. Spatio-kinematic information (radial velocity maps including peak radial velocities) was derived from the spectra every 0\farcs5 from the star with Gaussian fits after correction for uneven slit illumination (see Section \ref{kin}).

\section{Results}
\label{results}

\subsection{Preliminary inspection of the line ratios}

The distribution of the line ratios together with their errors allows us to estimate the data quality of the input values. This, in turn, gives an indication of the reliability of the derived physical parameters at each position from the source and suggests ways to improve the calculation.

\begin{figure}
\centering
\includegraphics[width=14.3cm]{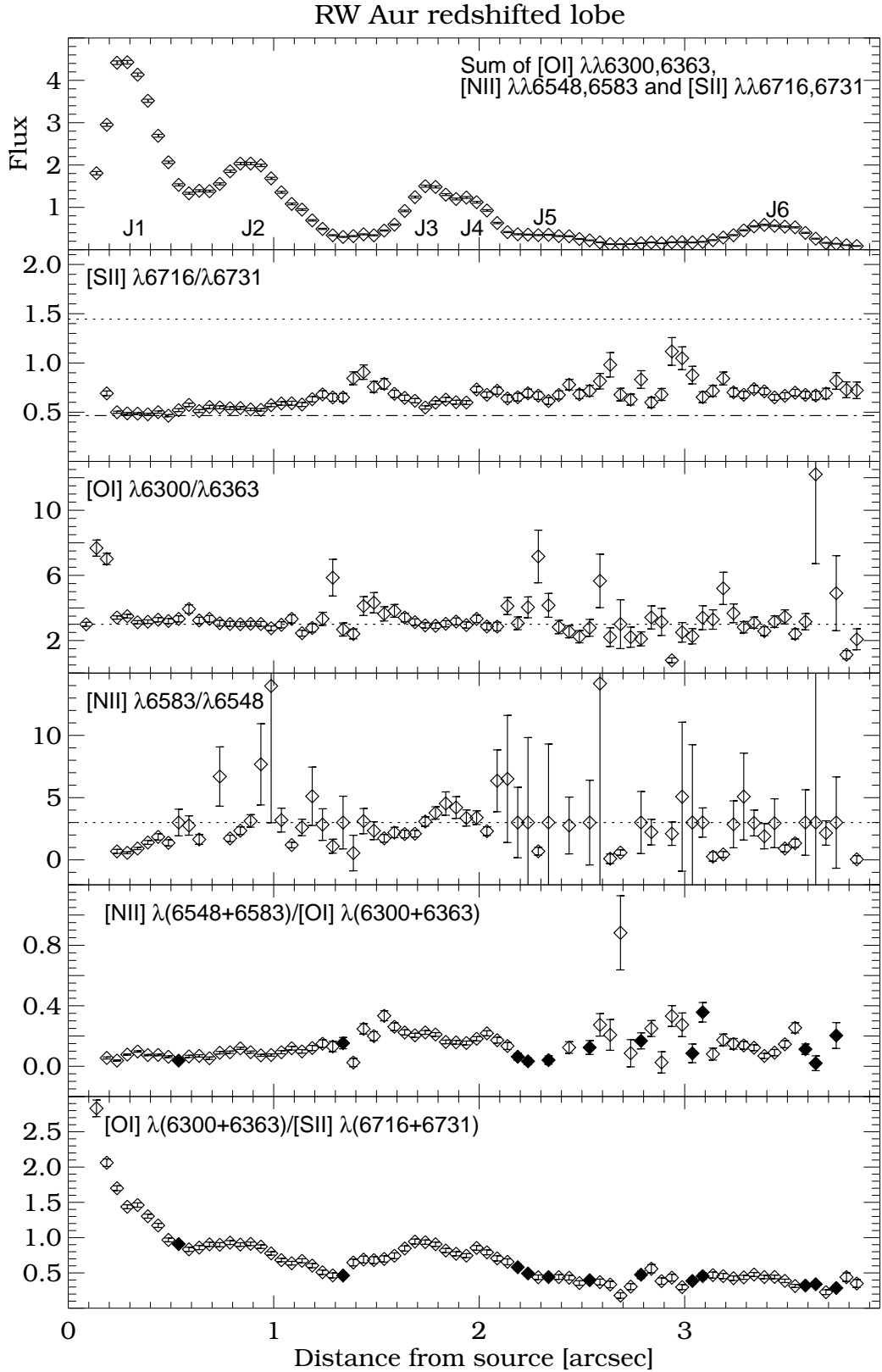}
\caption{\label{ratiored}
Ratios of the optical forbidden lines along the redshifted lobe of the bipolar jet from RW~Aur. From top to bottom: intensity profile resulting from the sum of forbidden lines integrated in velocity and across the jet width, in units of $10^{-14}$\,erg\,s$^{-1}$\,cm$^{-2}$\,arcsec$^{-1}$, and with labelled knots; [S\,II]\,$\lambda$6716/$\lambda$6731 ratio and its error, with superimposed theoretical values in the low (dotted line) and high (dot-dashed line) density limits, respectively, calculated for $T_\mathrm{e}= 10^4$\,K; [O\,I]\,$\lambda$6300/$\lambda$6363 with its error, compared with the theoretical ratio of Einstein's A coefficients (dotted line); same as above, for the [N\,II]\,$\lambda$6583/$\lambda$6548 ratio; [N\,II]\,$\lambda$(6548+6583)/[O\,I]\,$\lambda$(6300+6363), with its error; [O\,I]\,$\lambda$(6300+6363)/[S\,II]\,$\lambda$(6716+6731), with its error. The positions where corrupted values were corrected (see text) are marked by filled symbols.}
\end{figure}

\begin{figure}
\centering
\includegraphics[width=14cm]{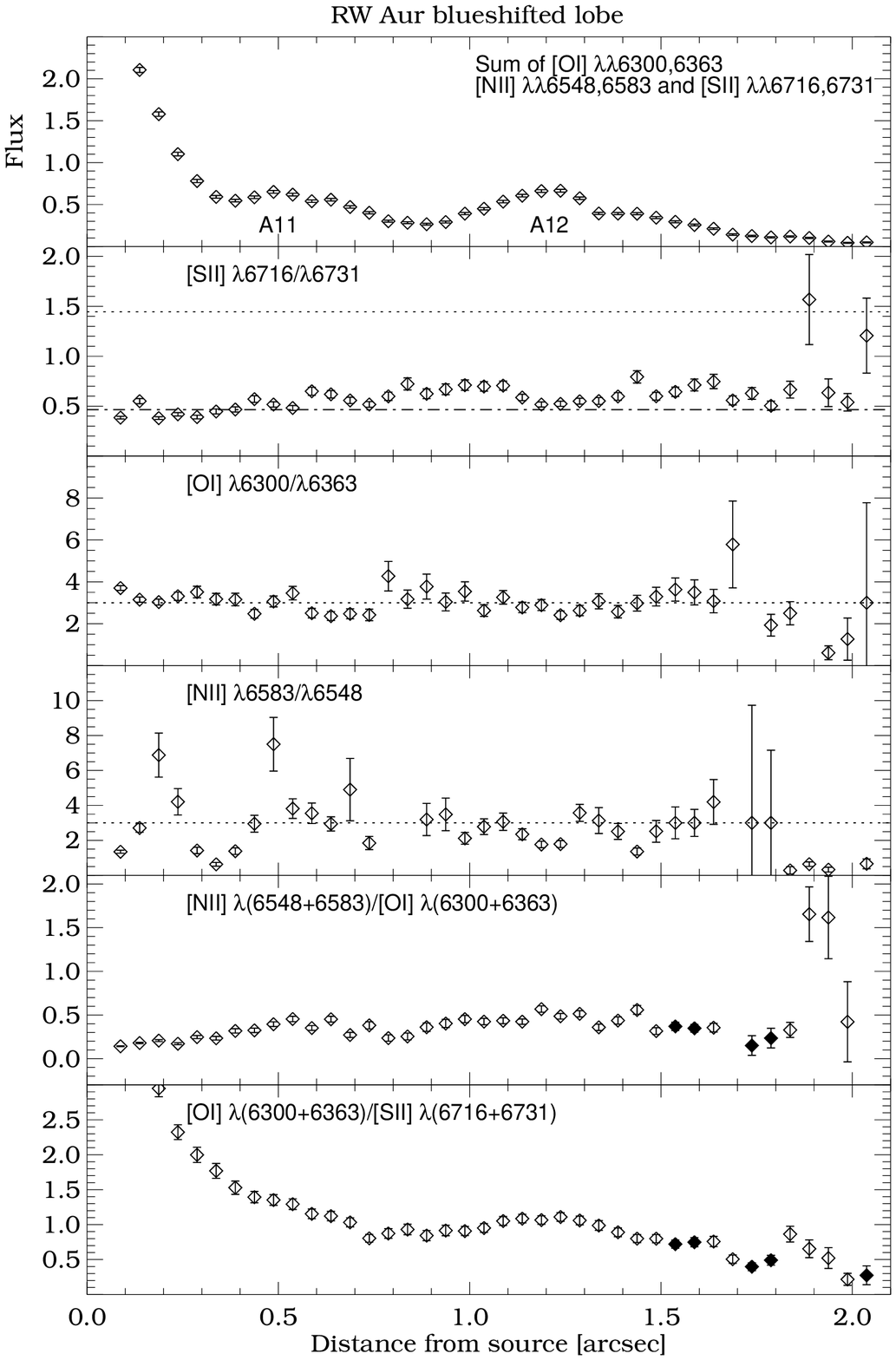}
\caption{\label{ratioblue} 
Same as Fig.~\ref{ratiored}, but for the blueshifted lobe of the RW~Aur jet.}
\end{figure}

In the top panels of Figs.~\ref{ratiored} and~\ref{ratioblue} we report the values of the line ratios [S\,II]\,$\lambda$6716/$\lambda$6731, [O\,I]\,$\lambda$6300/$\lambda$6363 and [N\,II]\,$\lambda$6583/$\lambda$6548 in the red and blue lobe, respectively, compared with the global intensity profile along the jet. The ratios and the profile are calculated from the lines extracted from the original spectra with a spatial sampling of 0\farcs05, and subsequently integrated in radial velocity and across the flow. In the plot for the [S\,II] ratio the limiting values for the $n_\mathrm{e}$ diagnostics of 0.465 (high density limit, corresponding to the critical density for collisional de-excitation) and 1.445 (low density limit) are shown. Values outside this range do not provide a good diagnostic determination for the electron density, and in the corresponding positions we have assumed that the electron density is equal to the critical density, 2.5$\times10^4$\,cm$^{-3}$, or 50\,cm$^{-3}$, respectively, thus giving a lower or upper limit to the actual electron density (see next Section).

In addition, the [O\,I]\,$\lambda$6300/$\lambda$6363 and [N\,II]\,$\lambda$6583/$\lambda$6548  ratios should be equal to the ratio of Einstein's A coefficients for spontaneous emission in each species, which is constant and equal to about 3 in both cases. In the positions where values of these ratios show consistent deviations from the theoretical values, either one or both lines of the doublet were weak and/or corrupted by imperfect subtraction of the stellar continuum or isolated hot pixels, or by the presence of strong H$\alpha$ emission. The situation is more critical in both lobes for the  [N\,II]\,$\lambda$6583/$\lambda$6548 ratio: in the first 0\farcs5 from the source this ratio is often exceedingly low, probably because of an underestimate of the [N\,II]\,$\lambda$6583 line flux caused by oversubtraction of the red wing of the H$\alpha$ line, that is very strong close to the source. Farther from the star, and especially after 2\arcsec, the situation is further  complicated by  the faintness of the [N\,II]\,$\lambda$6548 line, that, after subtraction of the H$\alpha$ halo, results in recognizable regions of artificial brightening or depression of the line, leading to uncertain and, in some cases, even negative values of the flux. The negative values are clearly unphysical and have been either dropped or corrected setting [N\,II]\,$\lambda$6548 = 1/3 [N\,II]\,$\lambda$6583  in the corresponding positions (identified by filled symbols in Figs.\,\ref{ratiored} and \ref{ratioblue}). In the other locations where the departures are observed it is difficult to establish which of the two lines is corrupted, so we preferred to keep the measured  values, using the panels of Figs.\,\ref{ratiored} and \ref{ratioblue} as a flag map to defective positions.

\begin{figure}
\resizebox{\hsize}{!}{\includegraphics{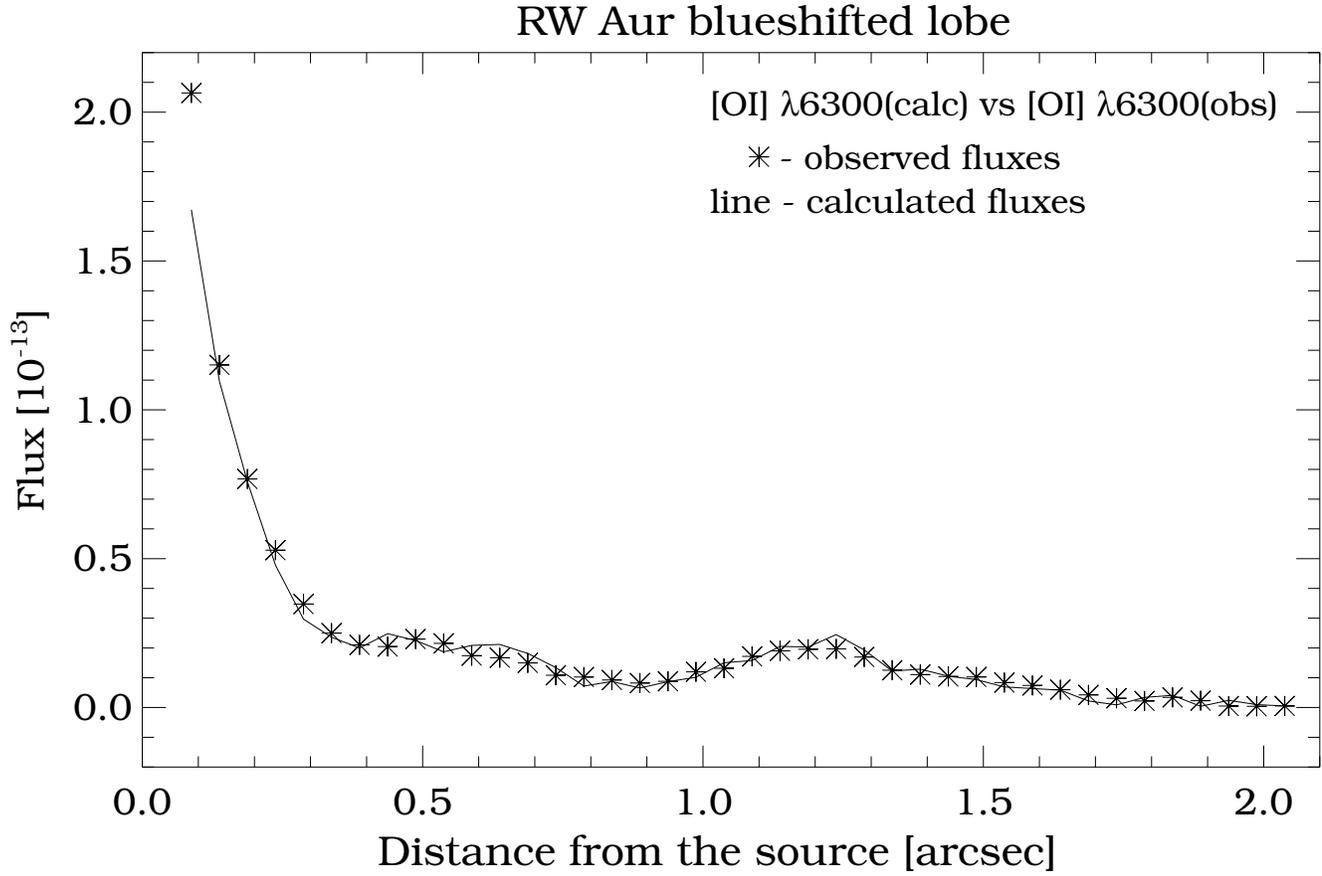}}
\caption{\label{obs_vs_art} 
Calculated values of the integrated flux in  [O\,I]\,$\lambda$6300 (\textit{asterisks}) computed as Flux([O\,I]\,$\lambda$6300)$_\mathrm{calc}$ = Flux([O\,I]\,$\lambda$6363)$_\mathrm{obs}\times3$ according to the theoretical ratio of Einstein's A coefficients, are plotted together with the observed integrated fluxes of [O\,I]\,$\lambda$6300 (\textit{line}). This demonstrates that no significant flux is lost because of the [O\,I]\,$\lambda$6300 line lying at the left border of the detector.}
\end{figure}

We note also that in the blue lobe, the blueshifted [O\,I]\,$\lambda$6300 line lies  close to the edge of the CCD (see Fig.~\ref{rvmap}), and the emission on the left border is partly cut. Nevertheless, by comparing at each distance from the star the integrated flux of the [O\,I]\,$\lambda$6300 spectral image with the integrated flux from a  similar image obtained from multiplying by 3 the [O\,I]\,$\lambda$6363 (Figs.\,\ref{obs_vs_art}), we have checked carefully that the flux lost is negligible everywhere (not more than a 0.5\% of the integrated flux), which is also confirmed by the plot of the [O\,I]\,$\lambda$6300/$\lambda$6363 ratio.

Finally, in the bottom panels of Figs.~\ref{ratiored} and~\ref{ratioblue} we show the  ratios [N\,II]/[O\,I] and [O\,I]/[S\,II] that are used as input to the diagnostic code. In both lobes the [N\,II]/[O\,I] ratio presents a slight monotonic increase with distance from the source, while the [O\,I]/[S\,II] ratio shows a pronounced decrease from the positions close to the source toward the jet beam. The latter effect, already reported for HH\,30 \citep{Bac99b} and other HH flows \citep{Har95}, occurs in the high density regions close to the jet source, where the [S\,II] lines are quenched, while the [O\,I] lines are not affected, having higher critical density ($\sim 2\times10^6$ cm$^{-3}$). In both ratios, local enhancements are seen in the red lobe at the position of knots J3 and J4.

The scatter with respect to the general trend is larger in locations corresponding to minima of the emission flux, as. e.g., in the regions  between the knots (e.g. at 1\farcs3 and 2\farcs8 in the red lobe). Spatial binning over the original sampling of 0\farcs05/pixel would have increased the signal-to-noise in these portions, but we avoided its use in order not to loose the unique information coming from the high angular resolution of HST data.

Another critical region is the first 0\farcs2 close to the star, because of imperfections in the subtraction of the continuum: despite the line ratios have been analysed here, this zone is excluded from the subsequent diagnostic analysis. Similarly, in both lobes the faint final portions of the beams (after 3\farcs8  and 2\arcsec\ in the red and blue lobes, respectively) could not be adequately treated by the BE technique and are not reported in Figures\,\ref{ratiored}, \ref{ratioblue} and subsequent figures.

\subsection{Physical quantities along the flow}
\label{discussion}

The set of line ratios illustrated above constitutes the input for the spectral diagnostics that leads to the determination  of the physical quantities in the emitting gas. The calculation is  performed numerically using a code developed to apply the BE technique to large datasets \citep[see][]{Mel08}. The physical parameters derived for the two lobes of the RW~Aur jet are presented in  graphical form in Figures~\ref{paramred} and~\ref{paramblue}. The main features are described below, separately for each lobe. The detailed kinematics of the flow is then illustrated in Sect.~\ref{kin}. Globally, the implications of these results are discussed in Sect.~\ref{disc}.

\subsubsection{Red lobe}
\label{resred}

In Fig.~\ref{paramred} the top panel shows, for reference, the distribution of the coadded intensity  in the forbidden lines integrated in velocity and across the beam, along the first 4\arcsec\ from the source with a spatial sampling of 0\farcs05/pixel. The identified knots are labelled J1--J6, following the naming scheme of \citet{ME98} (the three brightest knots at the same jet section are labelled by \citet{Lop03} as R4--6). Similarly, we report the profile of the  peak radial velocity in [S\,II]\,$\lambda$6731 and [O\,I]\,$\lambda$6300 for reference. This quantity is illustrated in more detail in Fig.~\ref{radvel} and analysed in Sect.~\ref{veldist}.

The next panel presents the electron density, $n_\mathrm{e}$ with its propagated error from the [S\,II] lines. In the first 0\farcs5 from the star $n_\mathrm{e}$ reaches the highest values, just below the critical density for [S\,II] diagnostics, $n_\mathrm{e,cr}= 2.5\times10^4$\,cm$^{-3}$ (error bars without the upper/lower tick indicate that errors on the [S\,II] lines lead the ratio above the limiting value). Then $n_\mathrm{e}$ gradually decreases, with an average value of 3400\,cm$^{-3}$ along the beam. A local minimum is found at 1\farcs4 and pronounced scatter is visible between 2\farcs5 and 3\farcs1, coincident with minima of the line emission in the top panel.

The hydrogen ionisation fraction $x_\mathrm{e}$ (fourth panel) gradually increases along the flow, from about 0.02 to an average of 0.07, but it reaches a value as high as 0.2 at 1\farcs5, on the left edge of the J3 knot. Other local maxima are present at the position of knots J2, J4, and J6, while the scatter in the faint portion between J5 and J6 is probably due to the uncertainty introduced by weak [N\,II] emission. It is interesting that at knot J3 the maximum of $x_\mathrm{e}$ is actually at the upstream edge of the structure. Such behaviour has also been noted in the HH\,30 knots by \cite{Har07}.

The electron temperature $T_\mathrm{e}$ calculated from the BE technique is illustrated in the next panel. The base of the jet presents the highest temperature, at $\sim 1.5\times10^4$\,K. This gradually decreases to 9000\,K at 1\farcs3, but then it sharply rises back to an average of $\sim 1.5\times10^4$\,K at the position of knots J3 and J5 (between 1\farcs5 and 2\farcs2). In the following faint section of the jet the average $T_\mathrm{e}$ is quite moderate, $\sim 10^4$\,K, and progressively decays toward the end.

The second to last panel from top  illustrates the total hydrogen density $n_\mathrm{H}$, derived as $n_\mathrm{e}/x_\mathrm{e}$. The errors propagated from the other quantities are large in many positions, but on average one can recognize a slow decrease with distance from the source from $\log(n_\mathrm{H})\approx 5.7$ close to the source to $\sim 4$ at the last position examined. In the region of knots J3--J5, however, the density appears to locally follow the opposite trend, and likewise at knot J6.

The bottom panel of Fig.~\ref{paramred} shows the  mass outflow rate. The derivation will be discussed in detail in  Sect.~\ref{mdot} and using Fig.~\ref{mflux}.

\subsubsection{Blue lobe}
\label{resblue}

In this section we analyse the results for the blueshifted lobe, which is fainter than the redshifted one over the examined distances from the source (2\arcsec). The physical quantities derived in this region with the BE diagnostics are presented in Fig.~\ref{paramblue}.

The distribution of the radiative flux along the beam in the examined lines is shown in the top panel. In contrast to the red beam, the blue lobe does not present a prominent maximum close to the star, but it drops steadily from the first examined positions. The distribution of forbidden intensity is flatter, with less prominent knot features at 0\farcs5 and 1\farcs2, marked as A11 and A12 following the nomenclature of \citet{ME98} (or B3, B4 following \citealt{Lop03}). The radial velocity in the second panel from top is nearly constant at $-180$\,km\,s$^{-1}$ up to knot A12, and then it decreases to $-160$\,km\,s$^{-1}$ at 2\farcs0. Note, how velocities are much higher, in absolute terms, than the corresponding ones in the red lobe. The kinematic properties are presented in more detail in Sect.~\ref{kin}, while the velocity asymmetry is discussed in Sect.~\ref{disc}.

At the base of the lobe (first 0\farcs4 from the source), the electron density (third panel from top) is higher than in the red lobe, with  $n_\mathrm{e}>n_\mathrm{cr} $ in various positions (marked by arrows). Beyond this distance, the trend of the electron density is not uniform: $n_\mathrm{e}$ drops gradually, but there is also a local maximum at 1\farcs2, and a slight rise of $n_\mathrm{e}$ at the end of the section (at about 1\farcs8). Within the examined 2\farcs1  the average value of $\log(n_\mathrm{e})$ is 3.56, which is almost the same as in the red lobe ($\log(n_\mathrm{e})=3.61$) within similar distances from the source.

The ionisation fraction $x_\mathrm{e}$ (fourth panel) shows a substantial scatter along the way to 2\farcs1. Nevertheless, there is marginal evidence that $x_\mathrm{e}$ increases with distance from the source reaching local maxima of 0.3 and 0.4 at the positions of the knots. Such enhancement of $x_\mathrm{e}$ is also observed in the fainter lobes of other jets \citep{Bac99b,Bac02,Har07}. Beyond 1\farcs2, $x_\mathrm{e}$ appears to drop to 0.1, but  the errors are higher here. The average value $x_\mathrm{e}$ for the whole section is 0.23, considerably higher than in the redshifted jet ($x_\mathrm{e} = 0.08$).

The electron temperature $T_\mathrm{e}$ (fifth panel) is also higher than in the red lobe, being on average $T_\mathrm{e} = 1.6\times10^4$\,K over the studied section. Despite the large dispersion of the values, a slight increase can be identified in the first 0\farcs6, followed by a local minimum at 10$^4$\,K. After that, $T_\mathrm{e}$ varies around $1.6\times10^4$\,K which coincides, within the errors, with the mean value.

In the following panel one finds the total hydrogen density $n_\mathrm{H}$, calculated as $n_\mathrm{e}/x_\mathrm{e}$. This  shows a gradual decrease along the beam, with log($n_\mathrm{H}$) varying from 5.0 to 3.6. The arrows refer to positions in which $n_\mathrm{e}$ is above the critical density, and thus represent lower limits. In general, there are no large deviations from the mean value of $\log(n_\mathrm{H})=4.4$.

Finally, in the bottom panel of Fig.~\ref{paramblue} we report our determination of the  mass outflow rate. Again, its derivation will be discussed in Sect.~\ref{mdot}.

\subsection{Kinematic properties of the flow}
\label{kin}

The seven HST/STIS spectra of the RW~Aur jet, taken at high spatial and spectral resolution, provide the most detailed description of the kinematics of a bipolar collimated stellar jet close to its source to date. This information has been described in part in previous papers (\citetalias{Woi02}, and \citealt{Woi03,Woi05}). In this Section we complete the study providing an alternative description of the radial velocities along and across the jet lobes.

General indications come already from the position-velocity (PV) diagrams illustrated in Fig.~\ref{rvmap}. We recall that these isophotic maps, spanning the range from $-300$ to $+250$\,km\,s$^{-1}$, originate from coadding the spectra in the seven slit positions, and hence loosing the spatial resolution across the jet. In these maps the radial velocity  asymmetry is evident, as the material moves at about 100\,km\,s$^{-1}$ and $-190$\,km\,s$^{-1}$ in the red and blue lobe, respectively. The velocity, however, presents only little dispersion and variation along each lobe.

The knots do not present any identifiable double velocity component, but the velocity dispersion varies in the different lines. For example, the measured Full Width Half Maxima (\textit{FWHM}s) show that the jet has the lowest dispersion in [N\,II] and in the red lobe. Also, the dispersion is larger within knots than between them. On the other hand, the [S\,II] emission appears to be more uniform than the [O\,I] emission, while  [N\,II] is mostly concentrated at the knots. It is interesting that unlike the other forbidden lines, the flux in the [N\,II] doublet appears to be stronger in the blue lobe, testifying to its higher excitation (see Sect.~\ref{resblue}).

\subsubsection{Profile of the radial velocity along the flow}
\label{1Dkin}

In Fig.~\ref{radvel} we illustrate the variation of the peak radial velocity along the jet obtained from Gaussian fits to each emitted line separately, integrated across the jet width (that is, from the coadded seven spectra). The integrated lines have prominent profiles along most of the beam, which gives a good accuracy to the fit, (uncertainty is about $\pm 5$\,km\,s$^{-1}$, excluding the fit to [O\,I]\,$\lambda$6300 (see below), and except for a number of positions in which the [N\,II] lines are faint). The first two panels from top show a comparison of the radial velocities in the two lobes on the same scale, while the bottom panel illustrates the results for the entire red lobe. In general, the relative variation of the velocity is quite limited in both lobes, being less than 15--20\% of the average flow velocity all along the flow, and reaching its maximum value in the positions corresponding to knots J3--J4.

In the red beam (top and bottom panels) the trends in the radial velocity of the four bright lines [O\,I]\,$\lambda\lambda$6300,6363 and [S\,II]\,$\lambda\lambda$6716,6731 are similar, with little variation  around the average value of $+100$\,km\,s$^{-1}$ over the whole section. At the beginning of the jet the [O\,I] lines appear to have higher and more constant radial velocity than the [S\,II] lines, whose velocity, however, gradually increases to reach that of the  [O\,I] lines at about $<0\farcs6$ from the source. A possible interpretation for this difference may come from their higher critical density. Close to the source the [O\,I] lines trace preferentially denser material, probably located closer to the axis, while [S\,II] emission includes less dense material located towards the outer edges of the flow. This would suggest that the gas closer to the jet axis moves faster and may have been accelerated more than the gas at the edges. As mentioned above, the radial velocity correlates positively with intensity at knots J3--J4, being higher (about 120\,km\,s$^{-1}$) at these locations. This would be in line with the knots representing bow-shaped working surfaces in the flow. No clear correlation is observed, however, for the other knots. Despite the [N\,II] lines being fainter in our HST/STIS spectra, velocity determinations  were successful at a  number of positions. In general the variations of the [N\,II] radial velocities follow those of the other lines, and reach slightly higher velocities at the bright knots, consistent with [N\,II] emission tracing faster and highly excited material close to the axis \citep[cfr.][]{Bac00}.

In the second panel of Fig.~\ref{radvel} we report the values of  the radial velocity measured in the examined portion of the blue lobe. One notes again the asymmetry in the radial velocity of the two lobes, present in all lines, that makes the blue lobe faster than the red one by 1.5--2 times. This will be discussed in Sect.~\ref{disc}. Secondly, in the first arcsecond the [O\,I]\,$\lambda$6300 line presents a small shift with respect to the other forbidden lines, including [O\,I]\,$\lambda$6363. The shift is probably due to the fact that [O\,I]\,$\lambda$6300 lies close to the CCD border, which causes a lack of accurateness in the Gaussian fit. Values beyond 0\farcs7, however, are reliable and still show a slightly  higher velocity in the [O\,I] lines (as well as the [N\,II] lines) with respect to the [S\,II] lines, again probably related to the fact that [O\,I] and [N\,II] lines trace denser material closer to the axis. The gradual drop after 1\arcsec\, from the source correlates with the decrease in intensity. On the other hand, there is no correlation with the position of the knots. This behaviour is quite unusual and will be examined in the Discussion section.

\subsubsection{Spatial distribution of radial velocity.}
\label{veldist}

We have presented above the peak velocity profiles obtained from lines coadded from the seven spectra. From our dataset, however, one can resolve spatially the velocity field in two dimensions (along and across the jet). To this aim, spatial maps of the peak radial velocities have been constructed. First, to increase the S/N ratio we coadded in each spectrum the two components in each of the [O\,I], [N\,II], and [S\,II] doublets, with the exception of [O\,I]\,$\lambda$6363 in the blue lobe, as the other oxygen line lies on the border of the CCD and its Gaussian fit proved to be critical. Then for each spectrum, i.e. each slit position, and at each distance from the star we performed a Gaussian fit of the spectral profiles and found the peak velocity. We adopted a lower limit in the peak S/N ratio of 3 for the Gaussian fit to be accurate. Those locations in which the Gaussian fit proved difficult have been set to zero as for many positions in the [N\,II] lines. As mentioned above, a necessary step is then the application of a correction for uneven slit illumination at every position, that varies between 2 and 8\,km\,s$^{-1}$ in our case. For a full discussion of this point and for details of the calculation see \cite{Bac02a} and \cite{Woi05}. Finally, the  obtained  velocity values  were aggregated in a 2D map, which provides the velocity peak for each doublet and for each position resolved by STIS {\em along and across} both lobes of the jet. The radial velocity maps constructed in this way from the [O\,I], [N\,II], and [S\,II] doublets are shown in Fig~\ref{rvmap1}\footnote{Colour version of the map is available in electronic form at http://www.edpsciences.org}. The maps present the same velocity asymmetry discussed in previous sections: the material in the blueshifted jet is in general twice as fast as in the redshifted lobe. The 2D maps also confirm that the [S\,II] (and also [N\,II]) velocities at the base of the redshifted flow are lower than in [O\,I]. Furthermore, the 2D maps allow us to confirm that in the approaching lobe the material with higher velocity tends to be located closer to the axis. Such peculiarities are discussed in Sect.~\ref{asymm}.

Finally, we note that a marked asymmetry is present in the [O\,I] and [S\,II] images with respect to the position of the jet axis, especially in the red lobe. In fact, the material in the upper portions of the maps (i.e. right of the axis looking from the tip of the red lobe toward the source) has slightly more redshifted velocities than the material in the lowermost portion of the jet. This velocity asymmetry is known in stellar jets, and in previous papers has been tentatively interpreted as  rotation of the jets around their axes \citep{Bac02a, Woi05, Cof04, Cof07, Chr08}. The radial velocity maps of Fig.~\ref{rvmap1} are just another way of highlighting this effect for the RW~Aur jet. The sense of the velocity asymmetry is consistent with that found in \citet{Woi05}, and in other  HST/STIS spectra of the same jet taken with the slit {\em perpendicular} to the jet axis \citep{Cof04}. We add that, as mentioned in the Introduction, RW~Aur is the only jet for which a  discrepancy has been found between the rotation sense determined for the jet and for the accretion disk. The cause of this discrepancy has not yet been found, and a discussion of it is beyond the scope of the present paper.

\subsection{Mass outflow rate}
\label{mdot}

In Paper I the average mass and momentum outflow rate $\dot M_\mathrm{j}$,  $\dot P_\mathrm{j}$, respectively, have been estimated in an approximate way. Here we present a new, more detailed determination of the $\dot M_\mathrm{j}$ distribution along both jet lobes, exploiting two different methods described in \citet{Bac99b} and \citet{Nis05}.

The first method combines the derived densities and radial velocities with estimates of the jet radius $r_\mathrm{j}$,  to get the mass outflow rate through $\dot M_\mathrm{j} = \mu m_\mathrm{H} n_\mathrm{H} \pi r^2_\mathrm{j} v_\mathrm{j}$. In this expression $\mu = 1.41$  the average atomic weight per H atom \citep{All01} $m_\mathrm{H}$ is the proton mass, and $v_\mathrm{j}$ is the bulk spatial velocity. Since $n_\mathrm{H}$ is determined from  line ratios, this method is not affected by reddening (as the wavelengths are so close) or distance errors. Since ratios formed with different elements are used, however, there is a  dependence on the adopted elemental abundances. Furthermore, this method tends to overestimate the mass outflow rate because it assumes that the whole jet is uniformly filled at the derived total density, or, in other words, that the filling factor $ff$ is unity. This can be partially compensated for by the presence of regions in the flow even denser than those traced by the lines used in the BE technique. In any case, it should not be a significant factor affecting the RW Aur jet, as it appears to be very dense and collimated. The  bulk spatial velocities $v_\mathrm{j}$ at each distance from the star were calculated by de-projecting the radial velocities obtained previously with an inclination angle of the jet with respect to the line of sight of $i=46^\circ\pm3$ \citep{Lop03}. The radius of each jet section was assumed to be equal to one half the intrinsic \textit{FWHM} ($FWHM_{\rm in}$) of the jet brightness profile, determined from a jet image reconstructed from the seven spectra as in \citetalias{Woi02}, after coadding all forbidden lines and velocity intervals, and correcting for the \textit{FWHM} of the equivalent Point Spread Function (PSF) of the images, $FWHM_{\rm PSF}$.

The uncorrected observed profile width, $FWHM_{\rm obs}$ is completely analogous to the one obtained in \citetalias{Woi02} for the first 2 arcseconds of the jet on both lobes, and it follows the same jet opening angle in the  section from 2 to 4 arcseconds in the red lobe. Upon examining a reconstructed image of the stellar continuum, the $FWHM_{\rm PSF}$ turned out to be 0\farcs1, i.e. comparable to the standard resolution limit of the instrument. The intrinsic jet width was then obtained as $FWHM_{\rm in} = (FWHM^2_{\rm obs} - FWHM^2_{\rm PSF})^{1/2}$.  The correction is larger for positions close to the star where the flow diameter is small, while it  decreases with distance, becoming in practice negligible after 1\farcs5 from the star.

\begin{figure*}
\centering
\includegraphics[width=14.5cm]{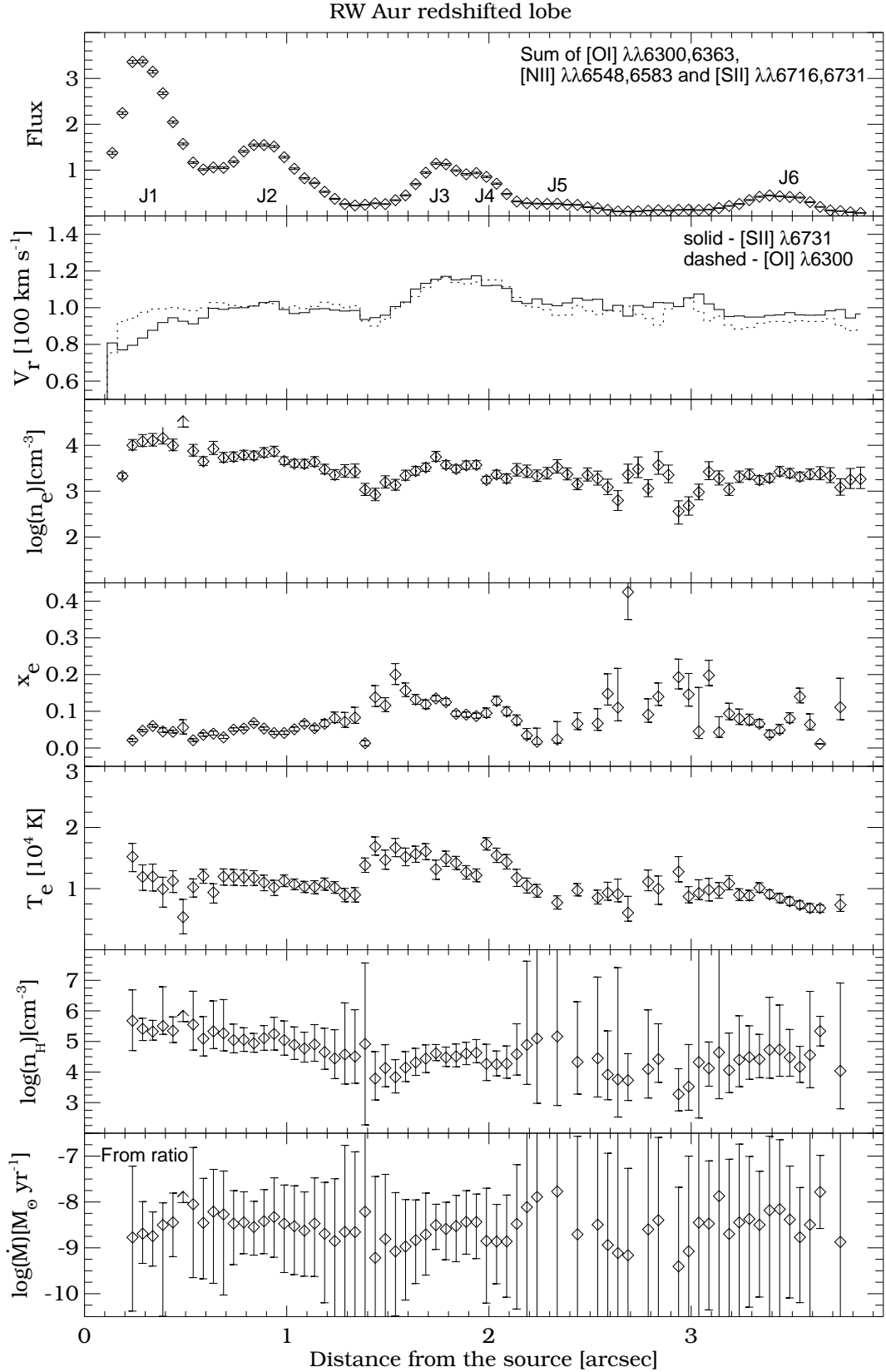}
\caption{\label{paramred} 
Physical properties of the emitting gas in the redshifted lobe of the RW~Aur jet as a function of distance from the source, derived from line ratios in Figs.~\ref{ratiored} and \ref{ratioblue} with the BE technique. From top to bottom: intensity profile as in Fig.~\ref{ratiored}; radial velocity in the [O\,I]\,$\lambda$6300 and [S\,II]\,$\lambda$6731 lines, integrated across the jet (uncertainty between 3 and 30\,km\,s$^{-1}$, see text); electron density $n_\mathrm{e}$; hydrogen ionisation fraction $x_\mathrm{e}$; electron temperature $T_\mathrm{e}$; total density $n_\mathrm{H} = n_\mathrm{e}/x_\mathrm{e}$ and mass outflow rate $\dot{M}_\mathrm{j}$ calculated from the above total density with the BE technique (see Section \ref{mdot} and asterisks in Fig. \ref{mflux}).}
\end{figure*}

\begin{figure*}
\centering
\includegraphics[width=14.5cm]{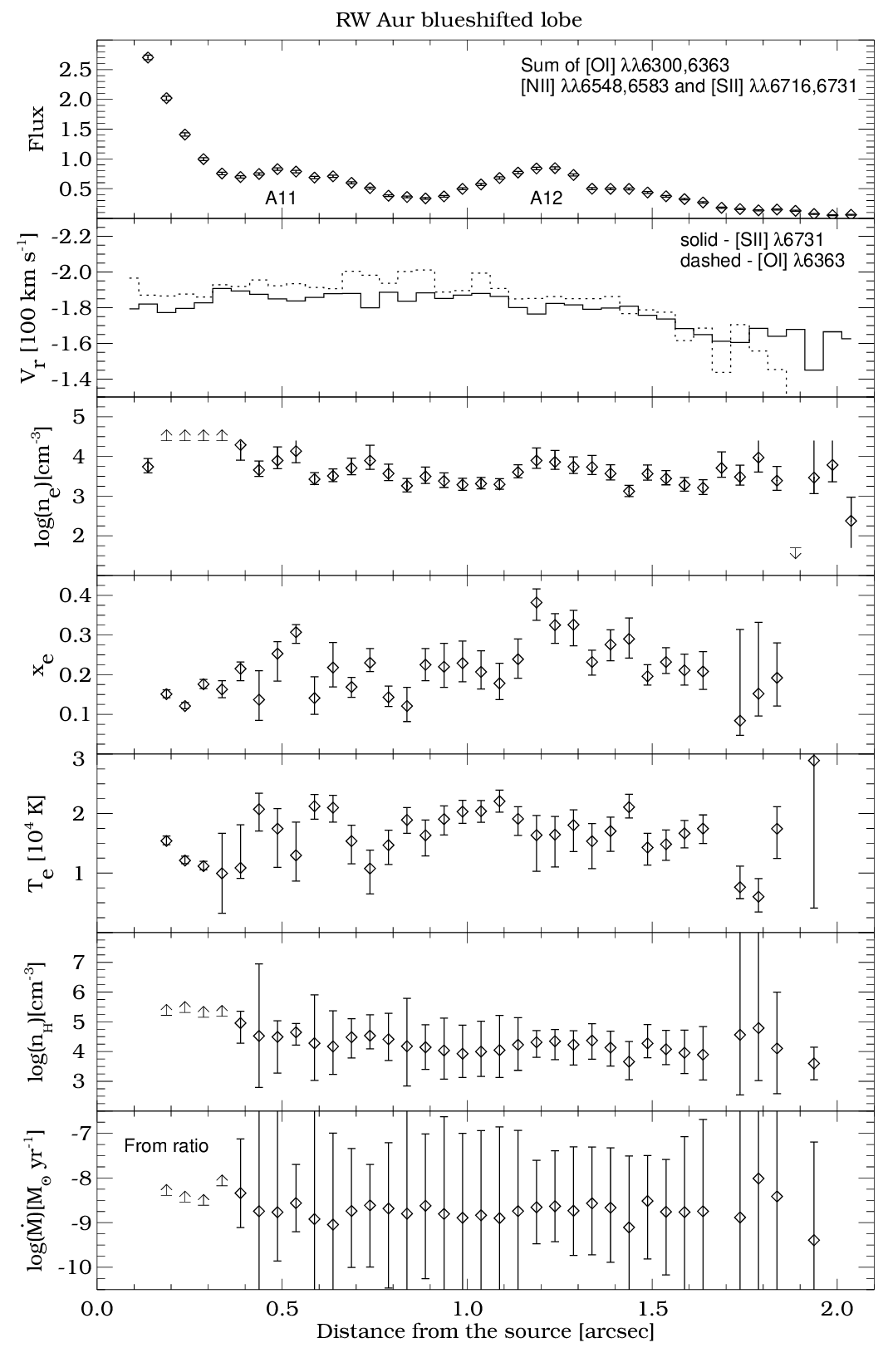}
\caption{\label{paramblue} 
Same as  Fig.~\ref{paramred}, but for the blueshifted lobe of the RW~Aur jet.}
\end{figure*}

\begin{figure*}
\centering
\resizebox{16cm}{!}{\includegraphics{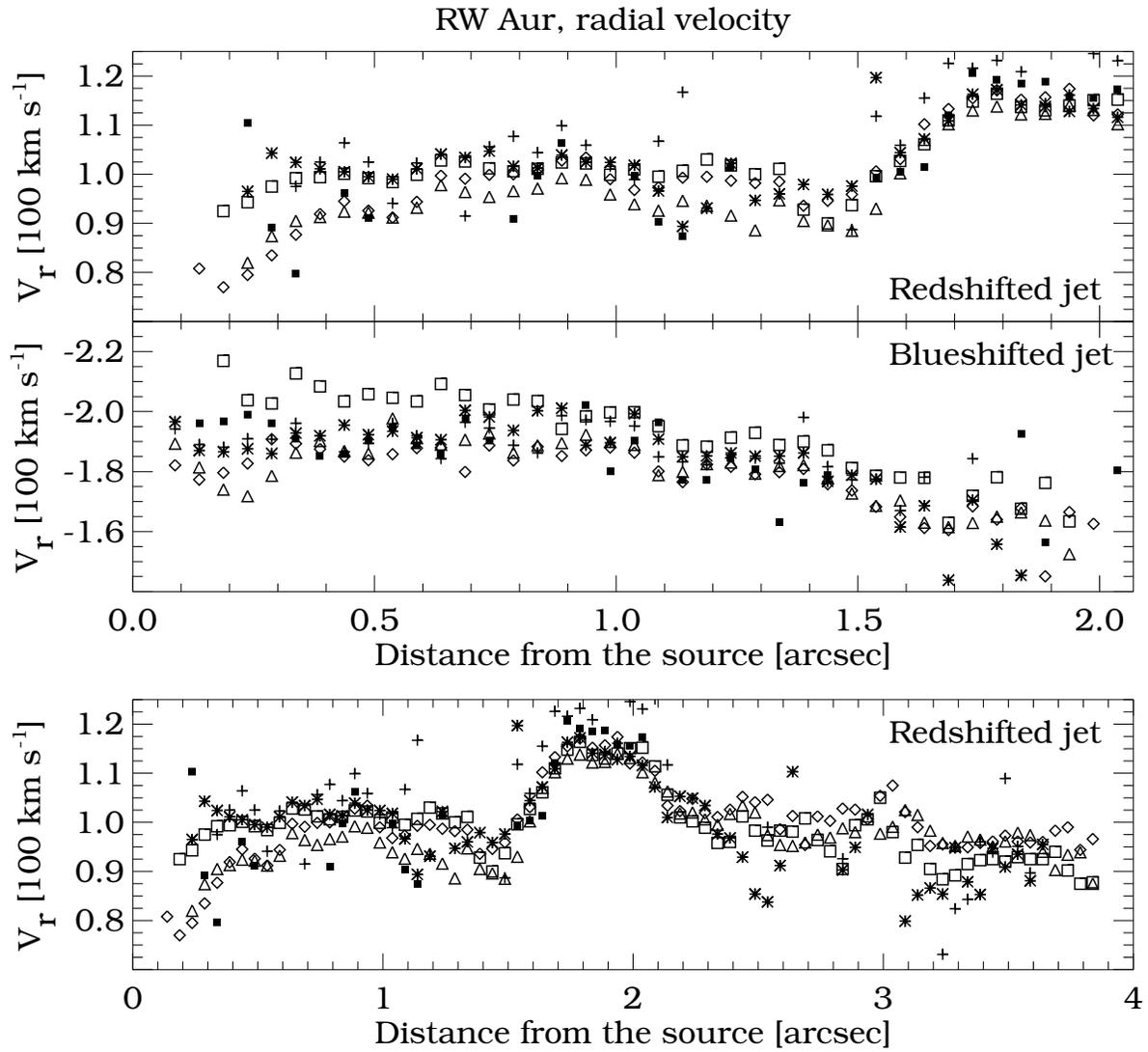}}
\caption{\label{radvel}
Radial velocity $v_\mathrm{r}$ measured in the  [O\,I]\,$\lambda\lambda$6300,6363, [S\,II]\,$\lambda\lambda$6716,6731, and [N\,II]\,$\lambda\lambda$6548,6583 lines emitted by the RW~Aur jet, as a function of distance from the central source. \textit{Top panels: } $v_\mathrm{r}$ distribution in the first  (2\farcs1) of the two lobes.  \textit{Bottom panel: } $v_\mathrm{r}$ for the whole red lobe section (3\farcs9) analysed. [O\,I]\,$\lambda$6300 is marked by empty squares, [O\,I]\,$\lambda$6363 -- asterisks, [S\,II]\,$\lambda$6716 -- triangles, [S\,II]\,$\lambda$6731 -- diamonds, [N\,II]\,$\lambda$6548 -- small filled squares and [N\,II]\,$\lambda$6583 -- pluses.}
\end{figure*}

\begin{figure*}[t]
\centering
\resizebox{\hsize}{!}{\includegraphics{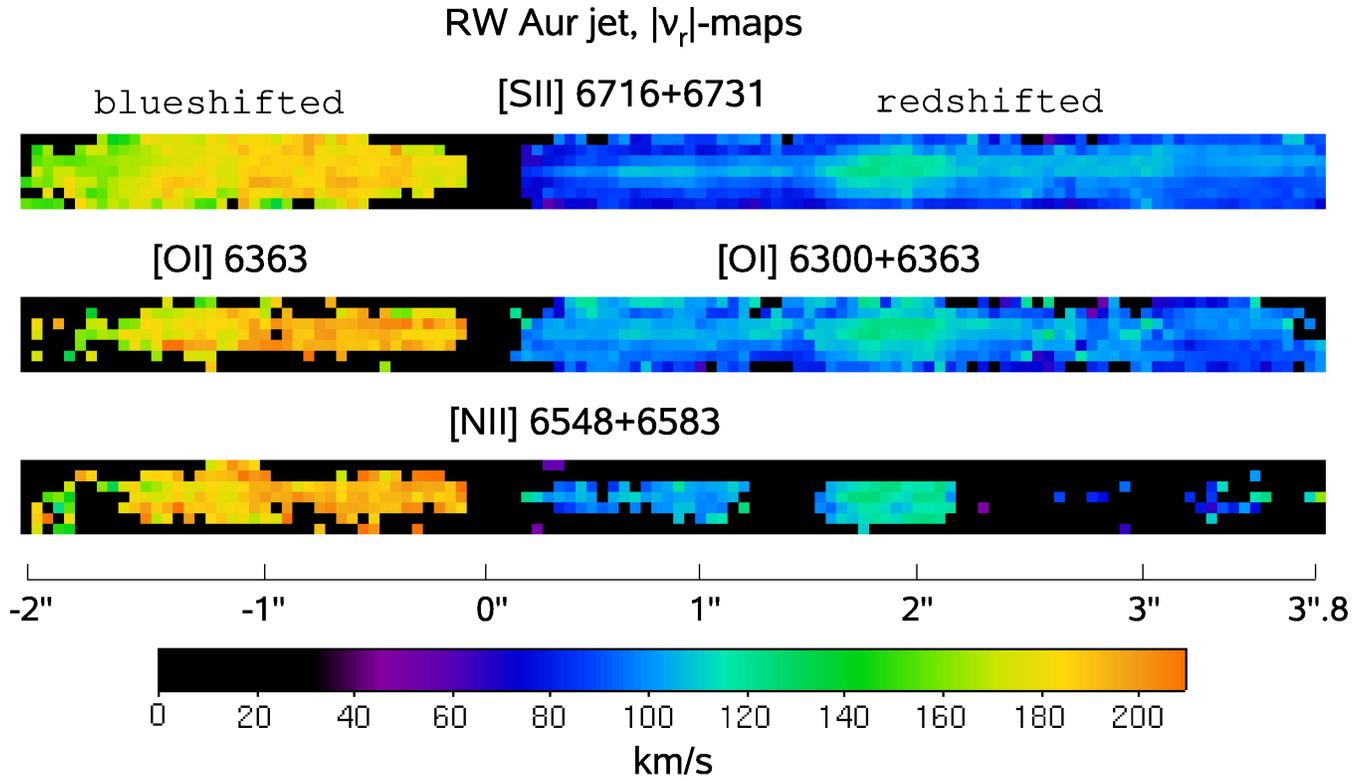}}
\caption{\label{rvmap1} 
Spatial distribution of peak absolute radial velocity $|v_\mathrm{r}|$ in the lobes of the RW~Aur jet, derived by aggregating the values obtained  by Gaussian fits to the line profiles in each of the seven spectra and at varying distance from the star. \textit{From top to bottom}: $|v_\mathrm{r}|$-maps obtained from the sum of the [S\,II] lines; same for the [O\,I] doublet (only the [O\,I]\,$\lambda$6363 line was considered in the blueshifted lobe because of the proximity of the [O\,I]\,$\lambda$6300 line to the CCD left edge); same as above, for the [N\,II] doublet. Positions where the Gaussian fit failed (because of low S/N) have their value set to zero. The colour scale for $|v_\mathrm{r}|$ ranges from 0 to 210\,km\,s$^{-1}$. All velocities are corrected for the heliocentric radial velocity of RW~Aur~A ($+23.5$\,km\,s$^{-1}$).}
\end{figure*}

\begin{figure}[t]
\resizebox{\hsize}{!}{\includegraphics{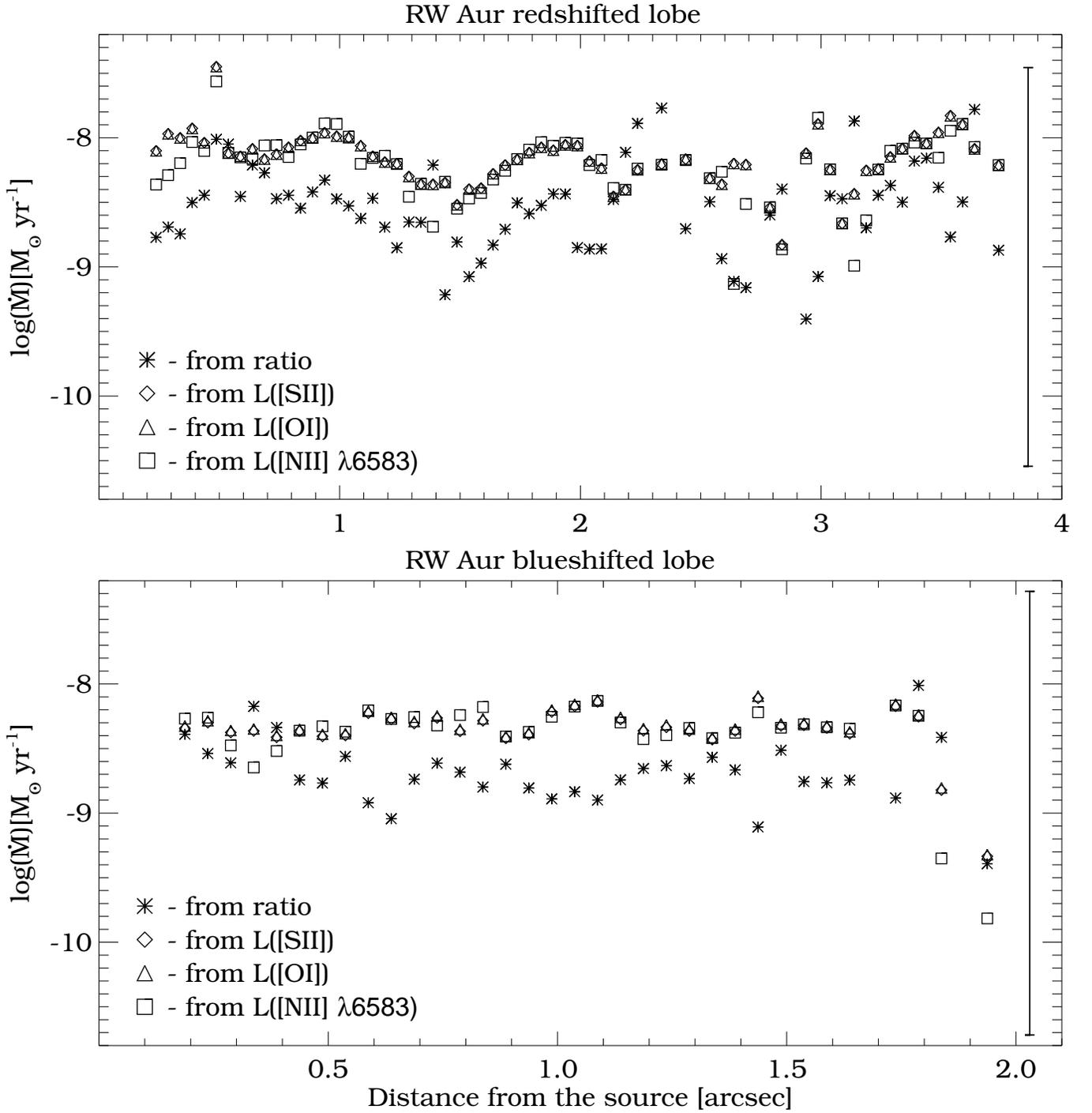}}
\caption{\label{mflux}
Comparison of mass outflow rates $\dot M_\mathrm{j}$ in the two lobes of the RW~Aur jet calculated from the total density obtained with the BE technique (method I, asterisks, labelled 'from ratio') and derived using absolute fluxes of forbidden lines (method II, diamonds: [S\,II]\,$\lambda$(6716+6731), triangles: [O\,I]\,$\lambda$(6300+6363), and squares: [N\,II]\,$\lambda$6583. The typical uncertainty on  $\dot M_\odot$ for each lobe is also showed with an average error bar at the right hand side of the panels. The averages correspond to $\Delta \log \dot M_\odot \sim 1.5$ for the red- and $\sim 1.7$ for the blueshifted lobe.}
\end{figure}

Combining these estimates with the total density $n_\mathrm{H}$ reported in Figs.~\ref{paramred} and~\ref{paramblue}, obtained with the abundance set of \cite{Est04}, one obtains $\dot M_\mathrm{j}$ in both lobes at each position from the star, as illustrated in the bottom panels of the same figures, and shown in larger detail by the asterisks in Fig.~\ref{mflux}. In Figs.~\ref{paramred} and~\ref{paramblue}, one can see that the error bars of $\dot M_\mathrm{j}$ are large in comparison with the range of this parameter. These errors are derived in an indirect way using $\dot M_\mathrm{j} = \mu m_\mathrm{H} n_\mathrm{H} \pi r^2_\mathrm{j} v_\mathrm{j}$ and combining errors for $n_\mathrm{H}$ and $v_\mathrm{j}$ into the mass flux errors according to the rules of error propagation. As a result, these error bars contain a systematic bias whereas the relative accuracy of the derived mass flux rates are much better and, in particular, the ratio of the rates in the red and blue lobes should be very reliable.

The second method is based on the fact that the observed forbidden lines are optically thin and thus their luminosity  is proportional to the mass of the emitting gas \citep[see][]{Har87}. In STIS spectra the lines are calibrated in brightness units, thus one can convert them into absolute flux and thus luminosity units and estimate the mass outflow rate by comparing the total observed luminosity in one line with the value calculated theoretically. In this case $\dot{M}_\mathrm{j}$ = $\mu$\,$m_\mathrm{H}$\, ($n_\mathrm{H}$\,$V$)\,$v_\mathrm{t}$/$l_\mathrm{t}$, with $n_\mathrm{H}\,V = L_\mathrm{obs}(line)/\epsilon_\mathrm{th}(line)$, where $V$ is the (unknown) volume effectively filled by the emitting gas, $v_\mathrm{t}$ is the tangential velocity of the emitting material, i.e. projected onto the plane of the sky, $l_\mathrm{t}$ is the length of the aperture considered, which we take equal to the extension of one pixel along the jet, $L_\mathrm{obs}(line)$ is the observed luminosity in the line in the same region and $\epsilon_\mathrm{th}(line)$ is the emissivity per particle for the line calculated theoretically as explained in, e.g., \cite{Pod06}. This method is affected by uncertainties in absolute calibrations, extinction, distance, and adopted elemental abundances, but does implicitly take into account the volume filling factor $ff$, which, in practice, is also simply the ratio of $\dot{M}_\mathrm{j}$ values derived from methods 2 and 1. For our calculation we used the total flux of the [S\,II] and [O\,I] doublets and the brightest [N\,II] line at 6583\,\AA. The adoption of the correct extinction value  is quite critical for this method, as it can sensibly affect the observed luminosity and hence the mass loss rate directly. A local determination of $A_\mathrm{v}$ with sufficient resolution is not available for this jet, and we have verified that this can introduce serious biases. The evidence, however, that the red lobe is, unusually, much brighter then the approaching lobe suggests that the extinction is different on the two sides of the system, and greater for the blue lobe, as if it propagates into the the parent cloud. We have chosen therefore to apply to the two lobes the extremes of the range derived from \cite{Whi01}, that is, $A_\mathrm{v}= 0.03$ and 0.72 in the red and blue lobe, respectively.

The values for the observed luminosity have then to be further corrected for the partial overlap of the slit area in the adjacent positions. The slit width is in fact 0\farcs1, but the positions are stepped by 0\farcs07, which has the net effect of counting twice the intensity falling globally on an area of $\sim$ 0\farcs18 times the slit length. We correct for this by multiplying the observed intensity by a factor 0.743 in both lobes.

Finally, we take a  distance of the system of 140\,pc, in calculating $v_\mathrm{t}$, and $\epsilon_\mathrm{th}(line)$ is given as a by-product of  the routines in our spectral diagnostic code. The mass outflow rates  $\dot M_\mathrm{j}$ calculated in this way are shown on a logarithmic scale in Fig.~\ref{mflux}, that illustrates the result determined from the various line luminosities and the comparison with the values obtained with the BE technique (first method, labelled 'from ratio'). The (quite significant) uncertainties are similar for the two methods, and are reported in Figs.~\ref{paramred} and~\ref{paramblue}, and, as an average indicative value in Fig.~\ref{mflux} with an error bar on the right hand side of the panels ($\Delta$ log $\dot M_\odot \sim 1.5$ and 1.7 for the red and blue lobes, respectively).

Overall, the mass outflow rate calculated with both methods is rather low, in the range $10^{-9}$--$10^{-8}\,M_\odot$\,yr$^{-1}$, and is almost constant along the beam. This is expected for a strongly collimated outflow with no sideways dispersion as in this case. Note however the decrease in $\dot M_\mathrm{j}$ seen in the red lobe and corresponding to the faint region between knots J2 and J3.

More precisely, excluding evident outliers, and, in both lobes, excluding the region beyond 2\farcs1, that might be affected by stronger interaction with the environment, from the BE technique one derives on average $\dot M_\mathrm{j} \sim 2.6\times10^{-9}\,M_\odot$\,yr$^{-1}$ in the redshifted lobe and $2.0\times10^{-9}\,M_\odot$\,yr$^{-1}$ in the blueshifted beam.

The values estimated with the luminosity method show a similar distribution with distance from the star, but are higher by a factor 2--2.5 all along the jet, which indicates that probably an unidentified systematic effect comes into play. For example, the radius of the dynamically important material in the jet might be larger than 0.5 $FWHM$, as assumed for the 'ratio' method. Alternatively, the peculiar reconstruction of the jet images may introduce an unidentified geometrical effect in the calculation of the luminosity of the lines, possibly related  to the unknown value of the local extinction. Because of the dependence of the latter method on extinction, distance of the object, and abundance/ionisation of the single species, we prefer to adopt the values derived from the BE technique. In any case the difference between the results from the two methods is much smaller than the uncertainty of the determinations.

Overall, however, the results obtained regarding the mass outflow rates is very interesting, because it indicates that {\em the jet has similar mass outflow rate on the two sides of the system, despite the evident asymmetries in velocity and density. This appears to suggest that the engine driving the jet is constrained on the two sides of the disk, as expected from axisymmetric magneto-hydrodynamic launch models.}

Further implications of this finding are discussed in the next Section. Here we just note that since approximately the same results are found with both methods, this suggests that indeed the filling factor is almost unity in this dense flow.  As mentioned above, however, this statement should be confirmed after a proper determination of the {\em local extinction} at each position, not available for the present study.

\section{Discussion}
\label{disc}

In this paper we have derived with the BE diagnostic method the set of physical parameters characterising the emitting gas in both lobes of the RW~Aur jet, with the best spatial resolution obtained so far along this flow. These are illustrated in Figs.~\ref{paramred} to~\ref{mflux}. To ease the discussion and compare with previous estimates, we also give a summary of the obtained parameters in the form of average values within the first 2\farcs1 along the jet in Table~\ref{table_param}.

In this Section we will discuss various aspects of the overall behaviour found in the derived quantities, starting from a comparison of the properties in the two lobes, in an effort to understand if differences in the ejection conditions may be present for the opposite sides.

\subsection{Comparison between the two lobes}

We consider the behaviour of the parameters in sections of equal length in the jet and counterjet, i.e. to distances of 2\farcs1 from the source. Note that this does not correspond to material ejected in the same elapsed time as we observe very different speeds on opposite sides of the central source.

First of all, the electron density $n_\mathrm{e}$ shows similar values in both the red and blue lobes, as well as the same gradual decrease with distance from the star, with oscillations corresponding to the variations of the total emission flux along the jet. The gradual decrease is consistent with a slow recombination of the gas from an initial ionisation state, combined with the gradual opening geometry of the jet, as discussed, e.g., in \citetalias{Bac99a}. The local increase of the electron density at the knot position can be justified by the presence of shock fronts that generate locally a compression of the gas and an increase in the ionisation fraction \citep{Har07}.

In contrast, the ionisation fraction $x_\mathrm{e}$ differs on average  in the red and blue lobes. In the red lobe, $x_\mathrm{e}$ starts from very small values ($\sim0.02$) and reaches its peak value at  $\sim0.2$. The ionisation fraction in the blue lobe at the positions closest to the source is higher ($\sim 0.15$) and reaches a maximum value of $\sim 0.4$, which is two times higher than the peak value in the brighter red jet. The asymmetry of $x_\mathrm{e}$ in the two lobes has also been discovered in other bipolar jets \citep[e.g. HH\,30,][]{Bac99b}, where the fainter lobe similarly exhibits a larger excitation. The electron temperature $T_\mathrm{e}$ is also higher in the blue lobe on average, (see Table~\ref{table1}). This may indicate that the brightness of the jet lobes in this case depends at least in part on the excitation conditions within the flow, in addition to a peculiar extinction pattern (the approaching blue lobe appears to be less luminous). This may also be due to fact that the density is higher in the red lobe (see below). In both lobes one notes a gradual increase of $x_\mathrm{e}$ over the first arcsecond or so, and then a decrease: such behaviour has been observed in several other cases at the beginning of jet flows \citep[e.g. HH\,30 and DG~Tau,][]{Bac99b,Lav00,Bac02,Har07}. A related finding may be that in both lobes for positions close to the source the electron temperature $T_\mathrm{e}$ drops from $1.5\times10^4$\,K to about $10^4$\,K. This may give hints to the origin, still unknown, of the initial ionisation at the base of the jet. The behaviour of the temperature as a whole in the rest of the beam  does not present any clear trend. It oscillates between $10^4$ and $2\times10^4$\,K, showing local increases at the position of the knots, but not always of the same amount.

Finally, the derived hydrogen density $n_\mathrm{H}$ in both lobes varies mostly in the range between $10^4$ and $5\times10^5$\,cm$^{-3}$, and it seems to be higher in the red lobe than in the blue one -- especially in the first arcsecond where it is higher by nearly one order of magnitude. This is in part the reason why the red lobe is brighter.

It is worth noting that despite the higher density in the red lobe, the mass outflow rate $\dot{M}_\mathrm{j}$ is similar in both the red- and blueshifted jets. Note also that these rates were derived using the BE technique, which is independent of extinction uncertainties. The fact that both lobes have approximately the same mass outflow rate, despite the numerous asymmetries in other quantities, may be important in modelling the launch mechanism (see below).

\subsection{Comparison with previous estimates}

Preliminary estimates of some quantities for the RW~Aur jet observed with HST/STIS were presented in a previous work from our group \citepalias{Woi02}. A comparison of Fig.\,\ref{sevenslits} of that paper with results presented here shows that the electron densities are in good agreement, as well as the observed \textit{FWHM} uncorrected for the \textit{FWHM} of the equivalent PSF (Fig.~3 in \citetalias{Woi02}), even if our determination of this quantity comes from the coadded image in all forbidden lines instead of a single line. The mass outflow rates instead present discrepancies, which are discussed below.

Tentative determinations of the  excitation parameters along the RW~Aur jet have been attempted in two previous studies by \citet{Dou02} and \citet{Bac96}. \citet{Dou02} quantified the physical properties from jet images obtained in 1997 at 0\farcs2 angular resolution with the OASIS spectro-imager mounted at the Canada-France-Hawaii Telescope (CFHT). The authors derived the excitation conditions in the redshifted jet of RW~Aur starting around 0\farcs4 from the source using the same forbidden lines (except [N\,II]\,$\lambda$6548), and diagnostic routines and elemental abundances similar to the ones described by \citetalias{Bac99a}. The derived $n_\mathrm{e}$ between 0\farcs4 and 3\arcsec\ (Fig.~3 in \citealt{Dou02}) agrees with our results to within 20\%. In contrast, the ionisation fraction $x_\mathrm{e}$ is generally smaller by about a factor of 2 than the one calculated from our HST spectra, although similar behaviour with distance from the star is found in both papers. As a consequence the hydrogen density $n_\mathrm{H}$ is also about twice the value determined here. Such differences in $x_\mathrm{e}$ may come from the fact that in the present work a different abundance set has been assumed than in \citetalias{Bac99a} \citep[see the discussion in][]{Pod06}. Nevertheless, the values found for $T_\mathrm{e}$ in \citet{Dou02} agree well with our estimates. Both papers report in the red lobe an increase in $T_\mathrm{e}$ of up to 1.3--1.5$\times10^4$~K between 1\arcsec\ and 1\farcs5, followed by a drop down to 7000~K between 1\farcs5 and 2\arcsec.

The excitation parameters in the red lobe were also analysed by \citet{Bac96}, who used spectra with low spatial resolution (1\farcs3) obtained with the Cassegrain twin spectrograph of the 3.5-m telescope at Calar Alto Observatory in Spain in 1993. The parameters derived by \citet{Bac96} present large discrepancies with respect to both our study and \citet{Dou02}. \citet{Bac96} found the ionisation fraction of the redshifted lobe decreases from a high value, $x_\mathrm{e} = 0.25$, near the star to about 0.02 at 6--7\arcsec. Within this distance a roughly constant hydrogen density of about $10^4$\,cm$^{-3}$ and an extremely low temperature (typically 4500\,K) were found. The reason for the big discrepancies is that in that paper a version of the spectral diagnostic technique including the H$\alpha$ line was used. As demonstrated later in \citetalias{Bac99a}, including the permitted H$\alpha$ line can lead to incorrect determinations of the physical parameters, in particular  higher ionisation fractions, and consistently smaller, unphysical temperatures. Contamination by chromospheric or magnetospheric H$\alpha$ is also a cause of the high ionisation values found close to the source in \citet{Bac96}. For these reasons the inclusion of H$\alpha$ is now avoided in more recent diagnostic studies. The derived average mass outflow rate in \citet{Bac96}, based on the results found at 4\arcsec\, from the source, was $\dot M_\mathrm{j} \sim 5\times 10^{-8}\,M_\odot$\,yr$^{-1}$. This is larger by one order of magnitude than our values, basically because of the different jet radius adopted, $R_\mathrm{j} = 10^{15}$\,cm \citep{Bac96}.

Recently, a set of HST/STIS spectra of jets from T Tauri stars including RW~Aur, taken in 2002 close to the source with the slit perpendicular to the flow have been analysed with the BE diagnostic technique by \citet{Cof08}. Of the two spectra, taken at 0\farcs3 from the source, only the one for the red lobe presented enough signal in the [N\,II] lines to be diagnosed. This gave a Position (with respect to the jet axis) -- Velocity diagram in the physical quantities, illustrated in Fig. 15 and 16 of \citet{Cof08}. The average electron density is $2\times10^4$\,cm$^{-3}$, the ionisation fraction varies between 0.02 and 0.07, and the total density is around $3\times10^5$\,cm$^{-3}$. These values are consistent with those found in the present work at the same position from the star. The electron temperature, however, turns out to be only $5\times10^3$\,K \citep{Cof08} a factor of 2 lower than observed here. This might be due to a very low value of the [S\,II]/[O\,I] ratio, which introduces large uncertainties in the temperature determination. In the same paper, an estimate has been given of the mass outflow rate in the flow, that turns out to be $1.7\times10^{-8}\,M_\odot$\,yr$^{-1}$, twice as large as our estimate at the same position. The difference comes mainly from the fact that in \cite{Cof08} the \textit{FWHM} measured in [O\,I] was larger by a factor of 1.4 than measured by us because no correction for the \textit{FWHM} of the PSF was applied. Such a correction is necessary particulary close to the star (see Sect.~\ref{mdot}).

The mass outflow rate determined in the present study for the red lobe is also smaller by a factor 2--5 (depending on position along the jet) than the one determined in \citetalias{Woi02}, and in \citet{Har95} from the spatially unresolved [O\,I] line luminosity at the source. This discrepancy is due to the fact that in \citetalias{Woi02} the ionisation fraction was taken from the study of \citet{Dou02}, which, as mentioned above, reports smaller ionisation fractions, and hence higher total densities than ours. In addition, in the absence of any other determination, the same ionisation fraction was adopted for the blue lobe in \citetalias{Woi02}, while we found here that the average $x_\mathrm{e}$ in the approaching lobe is more than twice that in the red lobe, which reduces the derived total densities. In addition, no correction for the PSF width was applied in \citetalias{Woi02}, which explains their larger values of $\dot M_\mathrm{j}$ in both lobes close to the source.

The reported differences may also be due to time variability of the ejection properties of the source, as the epoch of the observations analysed in the various studies differ (the data were taken in 1993, 1997, 2000 and 2002). T Tauri stars are known to vary rapidly in luminosity, and new jet knots have been seen to arise from jet sources on timescales of one year or less.

Finally, it is interesting to estimate from our measurements the ratio of mass ejection to mass accretion rates  $\dot M_\mathrm{j}/\dot M_\mathrm{acc}$, and compare it with those for MHD models for jet launch. An accretion rate of $\dot M_\mathrm{acc} = 1.6\times10^{-6}\,M_\odot$\,yr$^{-1}$ was determined for RW~Aur by \citet{Har95} from measurements of the veiling of photospheric lines caused by the accretion shock. For the ejection rate, one should consider the sum of the mass outflow rates of both jet beams. Adopting our average mass flux rate one obtains $\dot M_\mathrm{j} = 4.6 \times 10^{-9}\,M_\odot$\,yr$^{-1}$, then  $\dot M_\mathrm{j}/\dot M_\mathrm{acc} \sim 0.003$, which is compatible with MHD models with large values of the magnetic lever arm \citep{Fer06}. Note however that \citet{Val93} also analysed the veiling in RW~Aur spectra and derived a lower accretion rate of $\dot M_\mathrm{acc}= 3.4\times10^{-7}\,M_\odot$\,yr$^{-1}$ which suggests that $\dot M_\mathrm{j}/\dot M_\mathrm{acc}$ may be 5 times higher. Non-stationary accretion could also be at play, as RW~Aur is known to have a highly variable spectrum, in which emission lines undergo strong and rapid changes \citep[e.g.][]{MG82,Sto00}. This has been ascribed to fast variations of the accretion rate.

\subsection{Comparison with other young stellar jets}
\label{comp}

The same set of physical parameters (including the total hydrogen density) have been derived with 0\farcs1 spatial resolution for the DG~Tau and HH\,30 jets, as well as for the jet from the Herbig Ae/Be star LkH$\alpha$\,233 from HST images and spectra. This allows us a direct comparison with the present results for the RW~Aur jet. According to the findings in \cite{Bac99b}, the total hydrogen density of HH\,30 averaged over the first 4\arcsec\ of its redshifted jet is about $3\times10^4$\,cm$^{-3}$, while for LkH$\alpha$\,233 it is $1.1\times10^4$\,cm$^{-3}$ within the first 2\arcsec\ \citep{Mel08}. Thus, these jets appear to have lower densities than the RW~Aur jet, for which the average is $10^5$\,cm$^{-3}$. In contrast, the DG~Tau jet reaches values as high as $10^6$\,cm$^{-3}$ at its base \citep{Bac02a}, similar to the first positions in the RW~Aur red lobe. Other parameters in these jets present similarities with RW~Aur. As in the RW~Aur red lobe, in both HH\,30 and DG~Tau \citep{Bac00,Bac02a,Lav00} the ionisation fraction is seen to increase in the initial jet section, reach a plateau at 1--2 arcsec from the source, and then decrease slowly following a recombination curve. This has been interpreted as the ionisation fraction being produced close to the source, and then being 'frozen' into the jet medium. \cite{Har07}, however, demonstrated through their analysis of 2-epoch STIS slitless spectra that in HH\,30 there are ionisation peaks that follow the movement of the knots. Thus at least part of the ionisation is produced and maintained locally at the internal shock fronts. This is consistent with what we observe in RW~Aur, where we find a correlation between the peaks of $n_\mathrm{e}$, $x_\mathrm{e}$ and $T_\mathrm{e}$ and the positions of the bright features. This is particularly evident at knots J2, J3, and J6 in the red lobe. Furthermore, as noted by \citet{Har07} in HH\,30, we find a tendency of the ionisation fraction to increase upstream of the knot, instead of at the knot itself. A similar effect is also seen in the region before the bright condensation A in the \object{HH\,46/47} jet \citepalias{Bac99a}.

Several HH jets have been investigated on larger scales using ground-based telescopes (\citetalias{Bac99a}; \citealt{Pod06,Nis05}). The hydrogen density of knots in these jets is generally found to be between 600 and $6\times10^4$\,cm$^{-3}$, that is one to three orders of magnitude lower than what we measure in the first arcseconds of the RW~Aur jet. This indicates not only that the RW~Aur jet is one of the densest measured so far, but also that the material in the jet tends to be diluted as it propagates to large scales, both by beam opening and because of sideways losses due to internal bow shocks.

\subsection{Proper motion of the knots}
\label{prop}

The distribution of forbidden line emission along the RW~Aur jet reveals several knots clearly visible in the `total flux-distance' diagram. Positions of these knots have also been reported in \citet{Lop03} and \citet{Pyo06}, who investigated the propagation of some bright knots along the jet. With this goal in mind, they compared ground-based observations taken in 1997--2002, with our HST data taken in Dec. 2000 and first published in \citetalias{Woi02}. \citet{Lop03} found that the mean proper motion of knots for the red- and blueshifted jets are 0\farcs16 and 0\farcs26\,yr$^{-1}$, respectively, and this agrees with the velocity asymmetry of the flows. \citet{Pyo06} traced also the knots in [Fe\,II]\,$\lambda 1.644\,\mu$m. Comparison of knot positions in the red jet in the IR [Fe\,II] line with those in optical forbidden emission derived from HST spectra shows that the knot at 0\farcs25 exists at the same position in all investigated data sets and so, \citet{Lop03} and \citet{Pyo06} suggest that the knot may be stationary. In the blue lobe, there is another knot visible in the [Fe\,II] at 0\farcs17 which \citet{Pyo06} suggest is also stationary. At the same time, the HST data do not reveal any maxima at those positions in the optical. Instead, the forbidden emission flux steadily decreases from the position closest to the central source. The lack of the emission maximum in the optical might be due to problems with the extraction of the continuum as the star is very bright in this spectral range.

Using the proper motion of 0\farcs16 yr$^{-1}$ proposed for the red jet \citep{Pyo06} we suggest that the [Fe\,II] maxima at 1\farcs2 and 2\farcs29 \citep{Pyo06} correspond to knot J2 (0\farcs9) and J4 (1\farcs95), respectively, allowing for the 2-year period between the observations. While the positions of [Fe\,II] knots in the blue jet at 0\farcs48 and 1\farcs25 coincide well with knots A11 (0\farcs5) and A12 (1\farcs2) from the HST data, this may be coincidental. Instead taking into account the higher proper motion in the blue lobe, the knot A11 at 0\farcs5 (which is quite wide) may correspond to the [Fe\,II] knot at 1\farcs25. \citet{Pyo06} came to the same conclusion that the shoulder around 1\farcs25 may correspond to A11 at 0\farcs5 \citep[= knot B4 from][]{Lop03} if the knot has a proper motion of 0\farcs34$\pm0.1$ yr$^{-1}$.

\subsection{Jet asymmetry}
\label{asymm}

One of the most peculiar aspects of the RW~Aur jet is that it shows  asymmetries in several respects. First of all, the jet lobes show considerable difference in their velocities, noted already in \citet{Hir94}. The average peak velocity of the lines in the blue lobe is about $-190$\,km\,s$^{-1}$, while a velocity of only $+100$\,km\,s$^{-1}$ is measured on the opposite side. This asymmetry persists close to the source, as confirmed by the first analysis of our STIS dataset in \citetalias{Woi02}. Here we can add that  the [O\,I] and [S\,II] lines exhibit a systemic shift of velocities visible at the beginning of the red lobe: the velocity in [S\,II] is 10--15\,km\,s$^{-1}$ lower than in [O\,I]. The fluctuations of radial velocity  along each lobe, however,  are less than 15--20\%, with the largest variation seen in the prominent knots J3--J4. In the red beam the velocities  scatter around $+100$\,km\,s$^{-1}$, whereas in the blue beam a very slight decrease can be noted.

Additional evidence of the velocity asymmetry comes from analysis of the forbidden lines in other spectral ranges. Radial velocities of the [Fe\,II]\,$\lambda 1.66\, \mu$m line of RW~Aur jet have been reported by \citet{Pyo06}. Unlike the optical forbidden lines, the radial velocities of the infrared [Fe\,II] in the redshifted jet steadily increase from about 100\,km\,s$^{-1}$ at the beginning of the lobe to about 140\,km\,s$^{-1}$ at 3\arcsec. Because [Fe\,II] traces denser regions of the jet than [S\,II], this implies that the denser material may have had a higher velocity in the past, or is accelerating, whereas the velocity of material traced by optical [S\,II] lines remains almost constant. This suggests that denser material is located closer to the axis, where higher velocities are predicted by theory. On the other hand this  effect is not seen in [O\,I], which should also trace higher densities. In the blue jet the [Fe\,II] velocities behave similarly to the optical forbidden lines: they present a gradual decrease with distance from the jet source in the same range of velocities ($-200$ -- $-170$\,km\,s$^{-1}$).

The ionisation fraction $x_\mathrm{e}$ also shows evident asymmetries, while the electron density and temperature are similar in both jet lobes (see Table~\ref{table_param}). In the  blueshifted lobe the average $x_\mathrm{e}$ is 0.23, while in the redshifted one it is 0.08. As a result the redshifted jet is found to be 2 times denser. On the other hand the redshifted jet is about 2 times slower and this makes the  mass outflow rates similar on the two sides.

To find an explanation for the observed asymmetries is an open problem. The fact that the mass outflow rate is about the same in both lobes suggests that the engine driving the jet has certain constraining symmetries on both sides of the disk. Therefore, the origin of the differences in density, velocity, and excitation may instead reside in an inhomogeneity of the ambient medium around the star or, in a magneto-centrifugal launching scenario, in a different configuration of the circumstellar magnetic field that shapes both the accretion funnels and the open field surfaces along which the jet is launched.

The question if such an asymmetric configuration can be generated and remain stable for a sufficiently long time needs to be addressed with detailed theoretical modelling. It is known that a number of young stars do show asymmetries in the magnetic field distribution on their surfaces. The magnetically generated cold spots that cover a considerable portion of the surface of weak-line T Tauri stars (WTTS) \citep[see, for example,][]{Gra98,Gra08} present Doppler images consistent with largely inhomogeneous distributions and different properties in opposite hemispheres \citep{Ric96}. Strong asymmetries in the distribution of spots have also been revealed in evolved late-types stars, e.g., \object{XX~Tri} \citep{Stra99}. At the same time, \citet{Gra08} from analysis of the periodic variability of WTTS, argue that in some cases extended spotted regions can be stable over a timescale of at least two decades \citep[see the WTTS \object{V410~Tau} in][]{Stra99}. Recently, a complex non-dipolar topology of the stellar magnetic field has been discovered in the two accreting T Tauri stars \object{V2129~Oph} and \object{BP~Tau}, through spectro-polarimetric observations \citep{Don07,Don08}. However, a connection between surface magnetic properties and large-scale circumstellar magnetic configuration seems unestablished for the time being. On the other hand, there are theoretical models that propose a large-scale asymmetrical topology for the circumstellar magnetic field \citep[see, e.g.,][]{Lon07} which may ultimately explain the asymmetry of jet properties.

\section{Conclusions}
\label{concl}

In this paper we analyse in detail the first few arcseconds of both lobes of the RW~Aur jet, using long-slit high-resolution HST/STIS spectra. The physical properties of the outflow are derived applying the so-called BE technique to the spectra, described, e.g., in \citetalias{Bac99a} and in \citet{Pod06}. Our results  can be summarised as follows:

\begin{enumerate}

\item The distribution of forbidden emission along  both outflows reveals several local maxima identified as jet knots. We see seven emission knots in the redshifted lobe within 3\farcs8 from the source, and at least two knots in the blueshifted lobe within 2\farcs1. The distribution of emission within the knots in the blue lobe appear to be flatter than in the redshifted section.

\item The electron density $n_\mathrm{e}$ in both outflows is close to the high density limit ($\log(n_\mathrm{e})_\mathrm{crit}=4.39$) in the region near the driving source and decreases gradually with distance from the star. The electron temperature $T_\mathrm{e}$ in the redshifted lobe decreases steeply within the first 0\farcs3 from $1.5\times10^4$\,K to $10^4$\,K and then varies around this value. The same trend is observed in the first few positions in the blueshifted lobe, but then $T_\mathrm{e}$ reaches about $10^4$\,K and subsequently shows higher values with a maximum of about $2\times10^4$\,K.

\item We found that the gas in the red lobe has an ionisation fraction $x_\mathrm{e}$ ranging between 0.02 and 0.2, with an average of 0.08; in the blue lobe this parameter is higher and varies between 0.1 and 0.4 (average 0.23). $x_\mathrm{e}$ increases in the first section of both lobes, reaches a maximum at 1--2 arcseconds from the source and then starts to decrease. This behaviour appears to be a common feature also for other jets from young stars. Our results for $x_\mathrm{e}$ for the red lobe agree within a factor of two with those of previous studies \citep{Dou02,Bac96}.

\item The hydrogen density $n_\mathrm{H}$ in the brighter redshifted lobe appears to be twice as high (mean value $n_\mathrm{H}\approx 6.5\times10^{4}$cm$^{-3}$) than in the fainter blueshifted section, where an average value of $n_\mathrm{H}\approx 2.3\times10^{4}$ cm$^{-3}$ is found. RW~Aur jet is found to be the second densest \citep[after DG~Tau, ][]{Lav00} of the T Tauri jets investigated so far.

\item The kinematical properties of the flow have been fully re-analysed at high resolution and illustrated in accurate graphic detail. We confirm that the asymmetry in radial velocity ($-190$\,km\,s$^{-1}$ and  $+100$\,km\,s$^{-1}$ in the blue and red lobes, respectively) persists in all emission lines  and down to the source, and that the flow presents indications of rotation around its axis, in the same sense as determined in \cite{Woi05}. The pattern of proper motions of the knots has been updated with the data presented here.

\item Strong variations of $n_\mathrm{e}, x_\mathrm{e}$ and $T_\mathrm{e}$ are observed close to knots J3--J4 in the red lobe and A12 in the blue lobe. The radial velocity variations are very limited and reach a maximum of 20\% of the average at knots J3--J4 in the red lobe. Variations of the physical parameters at the other knot positions are suggested but are much less prominent.

\item The average  mass flux $\dot M_\mathrm{j}$ calculated from the forbidden line ratios is similar in the first 2\farcs1 of both lobes: $2.6\times10^{-9}\,M_\odot$\,yr$^{-1}$ for the red- and $2.0\times10^{-9}\,M_\odot$\,yr$^{-1}$ for the blueshifted jet. The $\dot M_\mathrm{j}/\dot M_\mathrm{acc}$ ratio is rather low and we estimate it to be between  0.003 and 0.012, depending on the adopted values for the mass accretion rate. This ratio, however, can still be accommodated by magneto-hydrodynamic models having a large value of the lever arm. We note however that the ratio may also be influenced by non-stationary accretion.

\item The overall trend of the physical parameters along the first few arcseconds of the RW~Aur jet is similar to that of HH\,30 and DG~Tau \citep{Bac99b,Bac00, Bac02a, Har07}, which have been investigated with similar resolution. Analogies in the  properties of the RW~Aur, HH\,30, DG~Tau jets in regions close to central star can reflect analogies in the mechanisms operating in that region, suggesting the same engine is  accelerating the jets in the T Tauri stars with outflows.

\item Our study of the RW~Aur jet indicates for the first time that, despite the detected marked asymmetries in physical and kinematic properties between the two lobes, {\em the  mass outflow rates in the two lobes are similar. This appears to indicate that the central engine has constraining symmetries on both sides of the system, and that the observed asymmetries are probably due to different environmental conditions.} We discuss the likelihood that these are  related to a large-scale inhomogeneity in the distribution of the local magnetic fields.

\end{enumerate}

In conclusion, our HST/STIS study of the physical parameters of the bipolar jet from RW~Aur has highlighted a number of properties that appear to be common to many objects of the same class when observed close to their source. That said, it reveals that peculiar elements of this and other jets, like the observed asymmetries in the two opposite lobes, remain unexplained even after analysis conducted at sub-arcsecond resolution. Nevertheless, this work shows that an accurate investigation close to the source of dynamically relevant properties, like the mass outflow rate, that present no asymmetry, can constrain models. Obviously whether mass outflows rate is balanced in other asymmetrical bipolar jets should be investigated. The combined efforts of future theoretical studies, and soon-to-become available interferometric observations at milliarcsecond resolution are expected to give useful signposts to an understanding of these phenomena.

\begin{acknowledgements} 
We thank Linda Podio for providing us with her code for the jet mass flux calculation. J.E. and S.M. acknowledge support from the Deutsches Zentrum f\"ur Luft- und Raumfahrt under grant 50 OR 0009. This study was supported in part by the European Community's Marie Curie Research and Training Network JETSET (Jet Simulation, Experiments and Theory) under contract MRTN-CT-2004 005592.
\end{acknowledgements}

\bibliographystyle{aa}
\bibliography{paper}

\newpage
\begin{table}[h]
\centering
\caption[Table]{\label{table_param}
Average physical parameters within the first 2\farcs1 along the RW~Aur bipolar jet lobes obtained with the application of the BE technique.}

\setlength{\extrarowheight}{2pt}
\begin{tabular}{l@{\hspace{2mm}}r@{\hspace{2mm}}r}
\toprule
\toprule
Parameters      & redshifted & blueshifted \\
\midrule
$n_\mathrm{e}$ (cm$^{-3}$) & 3600 & 4100 \\
$v_\mathrm{r}$ ([S\,II],[O\,I]\,$\lambda$6363) (km s$^{-1}$) & 100 & $-180$ \\
$x_\mathrm{e}$             & 0.08  & 0.23 \\
$T_\mathrm{e}$ (K)         & 12\,300 & 16\,400 \\
$n_\mathrm{H}$ (cm$^{-3}$) & 65\,900 & 22\,900 \\
$\dot M_\mathrm{j}$ ($M_\odot$ yr$^{-1}$) & $2.6\times10^{-9}$ & $2.0\times10^{-9}$ \\
\bottomrule
\end{tabular}
\label{table1}
\end{table}

\newpage

\newpage

\newpage

\newpage

\newpage

\newpage

\newpage

\newpage

\newpage

\newpage

\end{document}